\documentclass[11pt,fancychapters]{report}
\usepackage[a4paper, total={6in, 8in}]{geometry}
\usepackage{listings}
\usepackage{color}
\usepackage{setspace}
\usepackage{acro}
\usepackage{amsmath}
\usepackage{amsthm}
\usepackage{graphicx}
\usepackage{geometry}
\usepackage{subcaption}
\usepackage{cancel}
\usepackage{tikz}
\usepackage{color}
\usepackage{dsfont}
\usepackage[frozencache,cachedir=.]{minted}
\usepackage{caption}
\usepackage{comment}
\usepackage{authblk}
\usepackage[backend=bibtex]{biblatex}
\usepackage{doi}
\usepackage{colortbl}
\usepackage{booktabs}
\usepackage{hyperref}

\usepackage{pythontex}
\usepackage{listings}

\newcounter{example}[section]
\newenvironment{example}[1][]{\refstepcounter{example}\par\medskip
   \noindent \textbf{Example~\theexample. #1} \rmfamily}{\medskip}

\newtheorem{theorem}{Theorem}[section]   
\newtheorem{exercise}[theorem]{Exercise}
\newtheorem{solution}[theorem]{Solution}
\theoremstyle{definition}
\newtheorem{definition}{Definition}[section]

\usetikzlibrary{calc,trees,positioning,arrows,chains,shapes.geometric,%
    decorations.pathreplacing,decorations.pathmorphing,shapes,%
    matrix,shapes.symbols}
\include{acronyms}
\hypersetup{
    colorlinks,
    citecolor=black,
    filecolor=black,
    linkcolor=black,
    urlcolor=black
}

\definecolor{DarkerGreen}{RGB}{0,179,45}

\include{python_highlighting}

\newtheorem{exmp}{Example}[section]

\newcommand{\splitatcommas}[1]{\begingroup\lccode`~=`, \lowercase{\endgroup
    \edef~{\mathchar\the\mathcode`, \penalty0 \noexpand\hspace{0pt plus 1em}}%
  }\mathcode`,="8000 #1%
  }

\tikzset{
>=stealth',
  punktchain/.style={
    rectangle, 
    rounded corners, 
    draw=black, very thick,
    text width=10em, 
    minimum height=3em, 
    text centered, 
    on chain},
  line/.style={draw, thick, <-},
  element/.style={
    tape,
    top color=white,
    bottom color=blue!50!black!60!,
    minimum width=8em,
    draw=blue!40!black!90, very thick,
    text width=10em, 
    minimum height=3.5em, 
    text centered, 
    on chain},
  every join/.style={->, thick,shorten >=1pt},
  decoration={brace},
  tuborg/.style={decorate},
  tubnode/.style={midway, right=2pt},
}

\usepackage{glossaries}

\makeglossaries

\newglossaryentry{CI}
{
    name=Continuous Integration, 
    description={ blah blah }   
}

\newglossaryentry{abstraction}
{
    name=abstraction,
    description={An abstraction layer or simply an abstraction of the software is defined by interfaces that implement one or more use cases.  What the interfaces provide are well defined but how they they are implemented is not important, hence the term abstraction layer or abstraction level. For example, a list will provide an insert and query interfaces, among others, but it is not interesting from the point of view on the use of the list interfaces if the list is implemented as an array or using pointers. }
}

\bibliography{main}

\begin{document}

\title{%
Quality Engineering for Agile and DevSecOps on the Cloud and Edge \\
}

\author[1]{Eitan Farchi}
\author[2]{Saritha Route}
\affil[1]{IBM Haifa Research Lab}
\affil[2]{IBM Consulting}
\date{2024}

\maketitle

\begin{abstract}
Today's software projects include enhancements, fixes, and patches that need to be delivered almost on a daily basis to clients. Weekly and daily releases are pretty much the norm and sit alongside larger feature upgrades and quarterly releases. Software delivery has to be more Agile now than ever before. Companies that were, in the past, experimenting with Agile based delivery models, are now looking to scale it to enterprise grade. This shifts the need from the ability to build and execute tests rapidly, to using different means, technologies and procedures to provide rapid and insightful validation sequences and tests to establish quality withing the manufacturing cycle. This book addresses the need of effectively embedding quality engineering throughout the Agile development cycle thus addressing the need for enterprise scale high quality Agile development.  The role of the quality engineer is expanded to help to keep the end-user in the loop.  In the new role the quality engineer serves as its proxy thus better realising the Agile and DevSecOps principles at industrial scale.
\end{abstract}

\pagenumbering{gobble}
\newpage
\pagenumbering{roman}
\tableofcontents
\newpage
\pagenumbering{arabic}

\chapter{Motivation}

The way software is constructed has gone through a series of dramatic changes over time. Software was initially developed just as code, without much documentation, and then later, for enterprise scale more formal methods were required.  Therefore, formal methods came into being and they defined the software engineering life cycle processes leading to the build out of the waterfall model.  The waterfall model was a document-heavy process that included the writing of specifications, design and even detailed design and pseudo code level descriptions. This was coupled with a rigorous process to ensure review and approval.  The documentation, and the review and approval process happened in sequence before the code was actually written, compiled and run and hence the name the "waterfall model" in software construction. 

The waterfall model itself saw modifications and tailoring to adapt to different types of projects and delivery cycles. But the core remained the same, to move forward sequentially with requirements, design and code documented and reviewed through rigid processes.  The focus on quality in the waterfall model was a focus on testing, and a sequence of test activities  - static testing of requirements, design and code (namely reviews) and testing of the functionality delivered (unit testing,functional  testing, integration testing and user testing). 


In contrast, the current software construction model carries little resemblance to the original waterfall model and its variants.  Agile, Continuous Integration (CI), DevOps and Test First are the current dominating trends in software construction.  They represent a paradigm shift from the original way software was constructed.  Essentially, current software construction relies on short development cycles that enable quick feedback from the customer which is one of the key Agile principles that reduces risk.  Automation through DevOps enables the short feedback cycles.  Very light user stories are the new documentation and test first, when followed, may play a similar role in defining what the system needs to do or not do.  The current development practices all support the quick time-to-market requirement which is the ruling driver in today's software construction.  So software components are built in parallel.  A very key trigger for the change in the software delivery model from a waterfall development model is the emergence and rapid adoption of the API economy. With the API based application development approach,  solution components and component solutions are available for integration, versus a need to develop a monolithic solution from scratch. This is materialised in the Software-as-a-Service paradigm and impacted by Infrastructure-as-a-Service. These elements bring to the forefront more challenges to understanding the quality of the final solution. 


Quality engineering is the practice of ensuring that the software is doing what the customer wants it to do, by establishing  definitive checks and balances and implementing best practices and patterns, throughout the development process.  In Agile this is considered part of the overall build process and is expected to be a logical outcome of the definition of the user stories. There is a fine line between building in quality versus building code against requirements. Borrowing from methods for the education of children, we distinguish between 'Constructive Thinking' and 'Critical Thinking' activities in the software development process see \href{https://en.wikipedia.org/wiki/Six_Thinking_Hats}{six thinking hats}.  Constructive activities are focused on building, and are expanding and are creative in nature.  While Critical Thinking or critiquing activities are focused on attempting to identify what will break and why something will not work and how to manage it even while it is being built.  We make the observation that both are needed throughout the development process but with a continuous focus on the latter, namely, the critical thinking activities, to help build the right solution.  This book considers critical thinking or critiquing activities in the software construction process to be synonymous with quality engineering.  It addresses the best practices and potential opportunities for improvement of critical thinking activities in the current practice of software construction and in determining how it will be done and by whom.

Considering the agile practice, while quality is everybody's responsibility, one often fails to find clear ownership of quality engineering in the agile squad.  This leads to quality issues, project delays and pile up of backlog items. As a result, we often see Agile projects exceeding budgets and falling into a trap of playing continuous catch-up, instead of building forward.  In chapter \ref{process}, we discuss how to best integrate quality engineering roles into the Agile development and DevOps process.


One of the most fundamental principles of software construction is the hiding of implementation details or \Gls{abstraction}.  Good software interfaces can be used by simply understanding their input and output relations while ignoring how the interfaces are implemented.  One way to view the software design process, regardless of whether you follow a waterfall, agile or any other software development practice, is based on the manner of realising requirements. In the process of realising requirements, the software requirements are decomposed into interfaces that are then used to create use cases that meet the requirements. For example, to order a book in an online book shop, the system may call several interfaces to realise the action; one to determine if the book is available and to reserve it, another to determine if the customer credit is valid and yet another to notify the delivery department to handle the delivery of the book.  All of these interfaces are combined to meet the requirement of a customer having the ability to order a book.  The decomposition process continues recursively as the credit validation step may then use several credit validation processes depending on the method of payment being used by the customer. This decomposition is a crucial component of Domain Driven Design. While this process is well understood in principle, in practice the decomposition process typically does not include validating it in the context of customer needs.  We address this issue in chapter \ref{decomposition}.  

While code reviews are a common critical thinking activity practiced by agile teams, there are other effective review activities, which if performed, may uncover potential quality issues even while the sprint use cases are being defined.  In \ref{reviews} we discuss how this occurs and point out how to best utilize review processes in the agile context.


The most significant critical thinking activity is testing. What do we mean when we say testing?  Do we mean bug hunting or system design validation or both?  Test first takes the position that testing is done in order to design and craft the system well.  In \ref{testFirst} we show how to integrate the two approaches to testing to get the best results and to engineer quality into the solution.


To achieve the integration of quality engineering into the development process, you will require advanced technology and novel approaches to help drive testing for quality from design through to deployment and business usage (in the wild). New advances in AI can bring in automation and drive efficiency of the quality engineering processes across an extended software engineering life cycle to a new level. This is tackled in \ref{AI}.


There is an accelerated  digitization of enterprises and movement to cloud which brings deployment quality,  and quality in usage patterns to the forefront.  With the scale that edge accessibility brings, the rate at which modern cloud based micro-services are used have grown exponentially. As a result, resilience and performance of enterprise services have become crucial to business success.  In chapter \ref{AI} we will take a look at the emerging resilience challenge and discuss how to tackle it.

We assume familiarity with principles of Agile and DevOps throughout the book, while focusing on the quality engineering and its integration into Agile delivery.



\chapter{Quality Engineering in agile}  
\label{process}
 
 Today's software projects includes enhancements, fixes, and patches that need to be delivered almost on a daily basis to clients. Weekly and daily releases are pretty much the norm and sit alongside  larger feature upgrades and quarterly releases. Software delivery has to be more Agile now than ever before. Companies that were, in the past, experimenting with Agile\footnote{Note that we assume basic familiarity with Agile in what follows.} based delivery models,  are now looking to scale it to enterprise grade. This shifts the need from the ability to build and execute tests rapidly, to using different means, technologies and procedures to provide rapid and insightful validation sequences and tests to establish quality of the product within the manufacturing cycle. 
 
 We looked up Wikipedia \href{https://en.wikipedia.org/wiki/Quality}{(see link for evolving definition over time.)} for a common perception of quality and quality engineering. We found Quality defined as follows:  In business, engineering, and manufacturing, quality – or high quality – has a pragmatic interpretation as the non-inferiority or superiority of something (goods or services); it is also defined as being suitable for the intended purpose (fit for purpose) while satisfying customer expectations. Quality is a perceptual, conditional, and somewhat subjective attribute that may be understood differently by different people at different times in different ways.

When applied to software engineering the above definition of quality remains relevant but two important needs come to the forefront as we want to make sure that the code is built to address the requirements as well as built in a manner that is structurally sound. So we need to  
\begin{enumerate}
\item remove the subjective understanding of quality and
\item deliver a universally acceptable and reliable functioning code at speed. 
\end{enumerate}
The need for speed dictates the real-time / near real-time engineering of quality into code. Therein, emerges the construct of Quality Engineering in Agile. 

Whether you are already a tester and are looking to enhance your skill-set so you can embrace a role of driving quality engineering in Agile, or you are new to software development; the need to understand quality in Agile is paramount. The construct of quality engineering in Agile is fairly straightforward if you embrace the following definition.

\begin{definition}
IBM Consulting defines Quality Engineering as "the cultural mindset that leads organisations to embed quality principles and validations throughout the life-cycle, embracing available technology with the common goal of enabling predictability of business outcomes."
\end{definition}

Software Quality becomes everyone's responsibility irrespective of one's role in the team.  Role-based silos cease to exist as collaborative manufacturing extends from design to deployment and beyond and  includes all practices that enables engineering for quality.  Quality by design  requires a continuous focus on engineering to specifications and requires testing of the entire system and manufacturing process, as well as validating inputs and testing of intermediate artefacts and outputs.

We can realise this vision of quality-engineered software, with a set of concrete steps  and actions, that will allow you and everyone on the team to embrace the mindset of quality in the manufacturing life cycle. 
\begin{itemize}
\item Validate all artefacts (inputs and outputs) with the lens of an end-user as well as the lens of a consumer of intermediate artefacts.  (see \ref{decomposition} for a discussion of a definition of a user and a consumer).
\item Review and verify (inputs and outputs) so as to provide instant feedback by testing early and testing frequently against stated and unstated requirements.
\item Prevent defects instead of testing to uncover defects.  Drive quality with insights and value and not just metrics. Look at what can fail and what the impact of that failure would be.
\end{itemize}

These tenets are easy to adopt when  the definition of done for all outputs of the software and code produced includes a quantitative quality judgement.  Some examples of this follow.
 
\begin{itemize}
\item Do all requirements have quantitative acceptance criteria? (see \ref{reviews}
\item Was the code reviewed and did it pass the previously established review thresholds? e.g., only minor issues were found in the reviews.
\item Were all points of variation of the user story validated?
\item Was the code tested?
\item How many explicit tests were executed? 
\item How many tests passed?\footnote{Eitan - Mike - change to pass vs failed. Also some of the question marks have spaces.}
\item How many defects/blockers were identified?
\item How many validation points within the tests did the code pass?
\item How often/how many times were the tests executed? 
\item What was the test coverage?
\item Percentage of passed/failed tests
\end{itemize}

These checks can be included in daily scrum calls to become the de facto  completion criteria for any code that is checked-in and deemed ready for the build and deployment process. In addition, we validate the above against the acceptance criteria which covers both functional and non-functional aspects relevant to a service-oriented architecture. More details on designing quantitative acceptance criteria when we discuss Acceptance Test Driven Development (ATDD) in the next chapter.   

Note: The list above is indicative and you are advised to build your own lists as a team, to define the definition of done for each work product as recognised by the consumer. To do this right, you could conduct a Design Thinking workshop (see \href{https://www.ibm.com/design/thinking/}{LINK}.) or establish different ways to understand the needs and working models to be adopted. We have found Design Thinking to be a good model to bring stakeholders together to build in the quality needs for product and experience. The workshops will help the team to define QE in the context of the project and agree to include quantitative quality indicators into the definition of done for each deliverable and activity. 

The following exercise will help clarify the concept of building quantitative quality judgement for a definition of done.

\begin{exercise}
A test is typically a sequence of API calls.   For example, we may have implemented a function that checks if a given book is in stock, called $availablity()$, another function that checks if the customer has credit, called $credit_check()$, and last function that executes the order placement called $order_placememnt()$.  We have defined several tests needed for acceptance:
\begin{itemize}
\item  Tests that call $availablity()$ with different types of books, different number of copies and different types of availability
\item A sequence of checking availability and then checking credit only to find that the customer does not have the required credit.
\item A sequence that checks availability finds out that credit is available and then places the shipment.
\item A sequence that checks availability, finds out that credit is available but is not able to conduct the shipment to the desired destination. 
\end{itemize}
\footnote{tbc - 1. Maggie to create an AAT example/solution for this exercise. 2. Add to text - use a business flow view to create the tests.}

Explicitly implement the above tests and determine how many tests you have created, and how many points of validation each test contains.  Also determine the number of points of variation that the above story includes and how it relates to the tests you have created.  Do you think you have covered all points of variations? What additional functional and non-functional acceptance criteria would you add? 
     
\end{exercise}


\section{Driving agile delivery with QE}  

Quality Engineering, (QE), is a key lever to realising the Agile Manifesto. Integrating QE best practices into a modern Agile Development process, requires the right teaming (people and roles), and the right technology (methods and tools), to enable an efficient,  automated and insights-driven development and delivery. Different work-streams emerge as quality now moves beyond execution of tests to building the software such that is can be rapidly deployed to provide client value. So, while functional completion and correctness remain core and paramount, non-functional performance of the software, reliability engineering, and automation for reliability of the build and deployment process also comes into focus.

This means that designers,  developers and testers alike should see themselves as software engineers who are jointly engineer a quality component that can be deployed on its own to deliver client value.   This can be achieved by having a singular understanding of what defines quality (of requirements, of code, of tests, reliability of the software, of the DevSecOps pipeline, tools, and of completion / deployed software)  and hence ownership of the end product. For this, it is essential to develop the needed skills that allows one to perform in ones role efficiently. This may require one to build new skills for deconstructing requirements,  designing tests, embracing automation, scripting or automating tests as well as develop the ability to understand test results and gain  new insights from the same.  

Similar to the emergence of the role of the full-stack developer, there is value in including full-stack Quality Engineers in the project team. These engineers drive design thinking into the requirements refinement process, the design process including test design and embrace architectural principles into the code and the process of proving the functionality and features across the life cycle activities supporting the scrum lead and product owner. They drive quality product build-out against the definition of done with a maniacal focus on "designing for testablity" and "articulating acceptance criteria" including reliability and automation in DevSecOps in addition to functionality validation.

Some core skills that the Quality Engineers on the team  will need to build:

\begin{itemize}
\item Design Thinking 
\item Architectural Thinking 
\item Agile fundamentals
\item Model driven test and test optimisation methods
\item Test Driven Design
\item Behaviour Driven Development
\item DevSecOps fundamentals and 
\item Continuous Testing
\item Code and Automation integration
\item Feature flags and similar deployment constructs
\item Non-functional and chaos testing
\end{itemize}

 In addition, tool and automation/scripting skills to develop automated tests and include tests into the CI/CD pipeline has become essential. 

To help you get started on developing the above-required skills, you can access the Design Thinking Certificate course from IBM at \href{https://www.ibm.com/design/thinking/page/badges/core-skills}{link}) 

Remember, to truly establish quality of software systems, as a Quality Engineer, you also need to be aware of the underlying application technology and the deployment landscape. So you will need to improve your deployment stack knowledge and infrastructure skills such as cloud skills. Debugging, code review and scripting skills are also good to have. 

Further, as distributed cloud is the predominant deployment strategy, you will need to validate non-functional requirements as you test the systems performance thresholds by monitoring it at the micro-services level,  performing circuit-breaker checking, validating the logging, testing for resiliency and chaos management. These are a few independent QE and cross-over skills  that will hold you in good stead.

Note: Chaos Testing is an extreme implementation of critical thinking, as it looks to identify weaknesses in the application  - code, deployment and infrastructure by injecting adversarial actions / faults so as to bring down the system in a controlled yet chaotic manner against a planned hypothesis. This literally aims to crash-test the system at all planes.

Read more about chaos testing at - 
\href{https://www.ibm.com/garage/method/practices/manage/practice_chaotic_testing/}{link}).



\section{Driving enterprise transformation}

Enterprises are adopting digital operations with AI and cloud at a rate that is outpacing the manner in which traditional software development practices can support. Viewing Agile delivery as a mechanism to drive speed and address go-to-market needs, requires that  Quality is recognised as a journey where everyone participates, with a singular plan to deploy quality software for the business.  All participants in the journey are aware and understand the concepts of Agile, Quality, Testing and DevSecOps for continuous integration and deployment of the software.  All participants are aware that in the journey they may need to  embrace new ways of working and collaborate to craft quality into every client-facing and internal deliverable while potentially being open to learning new skills. A core skills and ask would be to elucidate all features and user stories with a clear description of what the final product should look like when viewed with a lens of what quality means in the context of the product being delivered. This will include elaborating the requirement or user story with a clear definition of done that is measurable in terms of the quality that the end user will experience.

\begin{exercise}
We need to determine if the following user story was implemented or not.  The user stony includes a search for an old favored book and suggestions of where to buy the desired used book.
Determine which of the following completeness criteria define done in the best way.
\begin{enumerate}
\item Code was implemented that searches for the favored book.
\item We have defined tests that searches for a book and determines whether it  exists or not and we will be done when all test pass.
\item We have defined tests that searches for a book that exists or not by one or more clients in parallel and we will be done when all tests pass.
\item We have defined tests that searches for a book that exists or not by one or more clients in parallel and we will be done when all tests pass and more than 80$\%$ of the code statements is executed.
\item We have defined tests that searches for a book that exists or not by one or more clients in parallel and we will be done when all tests pass, more than 80$\%$ of the code statements is executed and the solution also handles somehow the failure of a service.
\end{enumerate}
\end{exercise}

\begin{exercise}
Consider the following user story.  As a customer I wish to chose a pet and purchase it in the pet store rapidly.
\begin{itemize}
\item How can you improve the quality of the user story?   In other words, how can you make the user story more specific and complete?
\item How will you elaborate the definition of done and the acceptance criteria for this user story?
\item What functional and non-functional considerations will you include in the user story definition of done and the acceptance criteria?
\end{itemize}
\end{exercise}

Herein comes the product delivery mindset that is crucial to deconstruct user requirements so as to be able to build software and code that is complete, consumable and reliable. 
This can be achieved by a one team , one squad, one product mindset and requires the ability to continuously fine-tune and build the software while refining the understanding and quality of the product.  
To enable a successful product build the key elements include - understanding the business and user needs, putting the consumer at the center and building out the solution along with the customer. This means being able to design for the success of the user and business, and having the confidence to build right the first time around with the ability to enhance and add more features and capability.  The ability to change and enhance is pivotal and therein lies the importance of API-based delivery and the micro-services architecture in software development. It also underlines the need for QE in Agile.

\begin{exercise}
In the following scenario the architect of a solution talks with his friend and decides that the solution needs to have a new feature.   The architect then contacts the team of developers and brainstorming is conducted on user stories and a sprint defined.  A month after the initial idea was floated, the new feature is implemented.  Choose the most correct option below.
\begin{enumerate}
\item The above process indicates a well established Agile process that helps continually improve the solution.
\item Although the process of defining the requirements and the sprint are well followed actually this case indicates a completely non industrial mind set on the part of the architecture and the development team.
\end{enumerate}
\end{exercise}

In addition to a product delivery mind set, there are four elements for successful adoption of QE in Agile and for integration of QE in Agile delivery. 
\begin{itemize}
\item A mindset that enables engineering for quality to drive adoption of best practices, automation, proving and validation throughout the life cycle by all players in the team. Simply put it -  Quality is everyone's responsibility.
\item A potential redefinition and transformation of the way the work gets done, which leads to new responsibilities and skills. This includes embracing fluidity of roles performed by all players in the team. Simply put - Quality is achieved by everyone stepping up.
\item An agreement of what good looks like as seen with a business owners lens. Quality is cross-validated by measures, so as to indicate a successful business outcome. Which means that at every stage there has to be a proving of the business impact and value. Simply put -  the definition of done includes a quantified quality measure.
\item Establishing QE Operations Or QEOps in the continuous integration pipeline.  In other words making continuous proving against business needs an integral part of the engineering process.  Thus realizing automation and tooling that transcends roles and extends to managing quality activities as they are built and deployed. Simply put - Automate business validation across the life cycle
\end{itemize}




\section{Driving end-user value}


A product provides a business value to an end user.  For example, the end user may be a bank clerk.   The product helps the bank clerk make a decision whether to give a loan to a customer or not. A fundamental Agile principle is to put the user in the loop of the development process and get feedback in order to reduce the risk of developing something that the end user does not benefit from. How does one do this?  One naive way, would be to include the end user (designated stakeholder) on all project communications.  But that would not be wise as, there is more than one way in which this naive approach would break:

\begin{enumerate}
\item There are many clerks in a bank and there are many banks.  Which end user exactly are you going to put in the loop?
\item There are different end users the product needs to answer to.  For example, the bank itself requires that the overall business of giving loans is profitable and is less interested in the individual loan-giving decision as such.
\item Technically, even if an end user or a good stakeholder that can play its role could be identified, simply putting that stakeholder in the loop on all project communications is overwhelming.
\item There are different levels of abstraction in which communication on the product implementation is being conducted.  Specifically, a typical end user will not benefit from a description at a lower machine language level of the product.  
\end{enumerate}

To overcome these challenges, the Quality Engineer (QE) become a key stakeholder, functioning as an agent of the end user integrated throughout the Agile development process.  They don the mantle of the end user while being engaged continuously in the Agile development process itself.  They are key stakeholders in the process as they are skilled in clarifying the end users needs and in proving the requirements. They work to translate them into quality activities such as requirements validation, reviews and testing. These are performed throughout the development cycle; starting with and ending with acceptance tests at every stage and for every intermediary work product produced. These activities performed by the quality engineer serve as check points or quality gates, which can be shared with the client end user as identified by the business. An Olympic torch analogy can be used.  The end user is the first carrier of the torch, who passes the torch to the quality engineer to carry it forward through the development cycle, passing it back to the end user at relevant checkpoints.  Thus, the Agile delivery team benefits from the participation of the representative of the end user, who is continually involved in the creation, review, proving and validation of acceptance criteria and therefore of the software product itself as it is being built.  

In an industrial setting we cant really put the user in the loop, but what we can do is designate a member on the squad to be the user  or the quality engineer.

\begin{exercise}
\label{AgilewithQE:bizreview}

A developer describes the following design and would like to validate this as what the system needs to implement, with the end user.  The design is described in the following manner: $\forall x, y \in A f(x) \ne f(y)$ where $f()$ is a unique identifier of the book record.  How would you approach the challenge of describing this to the end user?   Select the best possible answer below.
\begin{itemize}
\item 
I will send the above design element to the appropriate stakeholder defined by the business for review.
\item
I would translate it to natural language as follows - "each book has a unique identifier " and then send it out for review with the stakeholders defined by the business.
\item
I will translate it to natural language as follows - "each book has a unique identifier".  Then create an example of a correct and an incorrect list of books and their associated identifiers.  I will then share it for review with the stakeholders defined by the business. 

\end{itemize}

See \ref{AgilewithQE:BiZreview:Solution} for the solution and explanation.
\end{exercise}

\begin{figure}[h!]
\centering
\includegraphics[scale=0.54]{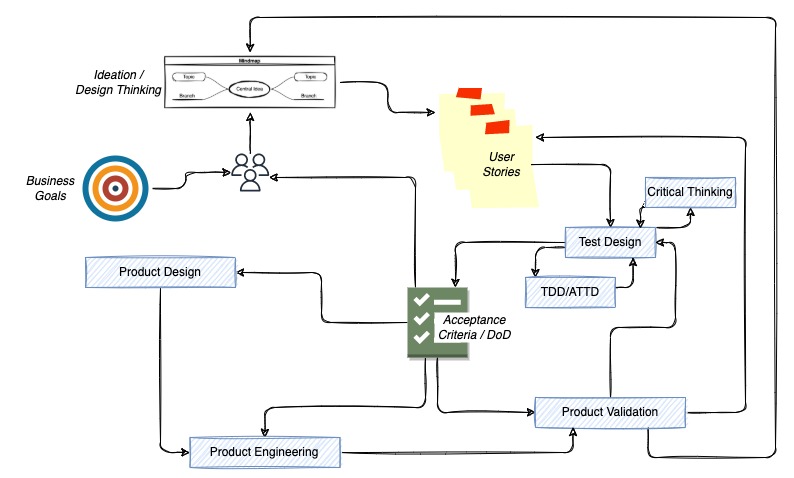}
\caption{How QE represents the end user}
\label{QE_user}
\end{figure}

In figure \ref{QE_user} We illustrate the manner in which business needs of the clients are kept in focus as we look to built out the software or product. This is done by putting Quality Engineering and test at the forefront and using the concepts of Testing as a central theme to build out the solution in line with the user's needs.

You will notice that the user requirements are established as goals / target state of the requirements that need to be realised by the product being built. The business needs (e.g., process a loan within 30 mins), becomes the focus of an ideation or designing thinking session. This will enable detailing of user stories or requirements. This could be a process of discovery and ideation run within the iteration.  The requirements when shaped are best validated for fitment to intended purpose and need by creating acceptance criteria or acceptance tests. The end-user is very much a part of this elaboration. Further, as the user stories for the product are detailed, and go into design and build, the same are taken in into a test design cycle. The outputs of the test design cycle feed back into the design and build process, before it loops back into testing and validation. You will see that the user or a QE who dons the user's persona is front and central in this validation process. We thus keep the focus on the business outcomes that the product is to deliver by leveraging the acceptance criteria at each stage of the build out.

The QE's role is to  examine the elaboration  using the lens of the end user and detail the acceptance criteria for each user story. These acceptance criteria include defining how the system will perform both functionally and non-functionally  - covering the user experience and the performance needs of the system.  These acceptance criteria are typically written in the language of the end user, describing the behaviour of the system as it is perceived and deployed. The acceptance criteria also becomes test cases that can be shared back with the business stakeholder / identified end user for validation.  In this manner, the sprint team gains clarity on what will be acceptable by the client and therefore builds what is acceptable. 

Throughout this process, there is a subtle change in the language of elucidation of the business needs into the product / into code. While at the earlier stages, the needs and the requirements will be more business-like and plain speaking language, as they go into the design and build process, the language nuances into a more technical and implementation construct. This could be in the form of Behaviour Driven Development (BDD) scripts, APIs, API specifications and automation code. This language would be typically beyond the ken of the end-user. Hence the QE becomes an essential conduit who uses the right language at the right stage and establishes the continuum of the solution from ideation to build in line with the requirements and needs of the specific user story in line with the larger goal of the product. Essentially, the QE serves as an interpreter without which a plausible communication gap could lead to a product that does not meet the business needs.  

It is interesting to note that often user stories get built but do not get delivered to production, primarily because of the above mentioned communication gap. This is often manifested in a piling up of features to be delivered, and then discarded as it becomes difficult to get customer sign-off because original buy-in was not established. 

This demands the tester learns new skills to embrace quality Engineering,  technical and soft-skills to bring the solution together and establish quality in the   product manufacturing process. We elaborate that in the next section.

As you work in an Agile team you will notice that the scrum leader also known as the team coach provides the required guidance to the team to encourage the build out of the user stories, and the quality engineer challenges the team on drafting the acceptance criteria. Thus both keep the focus on the goals and make sure that the product delivery is aligned to what the user has asked for and meets the requirements adequately.




\section{Skills and teaming for QE in agile}

In traditional software delivery the role of the tester is to focus on validating the end state of the software deliverable. The aim being to identify bugs.  In the product-based delivery model, the tester's role needs to be expanded to embrace quality engineering, as the need here, is to build the product in line with the business needs. Testing alone as an end state activity, will not achieve that. So the Quality Engineering practice should look at ways to not just prevent bugs but protect the build by aligning it continuously with the user stories and feeding that back into the design and build process itself. Over time we use tribal knowledge of working in the team, along with insights from defects and past history to predict quality pitfalls and prevent them.  As you look to transform from being a tester to a true quality engineering, you will need to hone the following capabilities over and above testing: 

\begin{itemize}
\item Agile and architectural skills
\item Domain and context awareness,
\item Automation and scripting skills,
\item Technical skills to enable integration of the work products at each stage to build it into the continuous integration and the DevSecOps pipeline.
\end{itemize}

Along with the capabilities of the Quality Engineer, we also see new roles emerging. These  include: 
\begin{itemize}
\item Transformation Consultants whose focus is to drive cultural change and enable the teams by coaching,  best practice inclusion within programs and across client leadership.

\item Product Quality Consultants - who will Coach and enable teams to embed quality and validation into their day-to-day delivery on their available technology stack

\item Product Quality Engineers - to embed alongside development engineers and digital workers alike, leading by example and using technology and automation to spread the load as the team grows to a point of quality being a team sport.

\end{itemize}

Another way to understand the role of the QE could be to consider levels of operating within the squad and models of working in Agile. We can consider the QE to operate at different levels  - doing, enabling and thinking.  So a product quality engineer could be 
\begin{itemize}
    \item scripting tests, integrating the automation, writing frameworks as a  performing role as a doer
    \item providing guidance, provisioning frameworks, enabling or configuring the pipelines and providing insights on the test results as a enabler 
    \item  engaging with the business stakeholders and the product owners to establish the key quality metrics and constructs and the rules of engagement to drive the definition and inclusion of quality and engineering best practices as a provocateur or thinker
\end{itemize}

It is not a formal hierarchy like that of the tester, the test lead and the test manager as seen in the waterfall testing world where the boundaries of operations and activities are expected to follow a regimented structure. Whereas in Agile, we expect the squad members to be self directed and  producing deliverable on their own, and taking on roles as needed to a produce a quality product.  However in large programs and to drive enterprise-wide adoption of QE in Agile, one might formalise the roles, though not as a hierarchy but as separate functions to be performed. This way we ensure that quality is driven forward to become everyone's agenda in the organisation in a coherent and cogent manner. this could require the inclusion of a Senior Quality Engineer or Architect who will guide and provide guidance across squads to embed the right quality engineering principles. This role works in tandem with the Product owner, the Team coach and the Agile coach to build in the culture and practices of quality. This role becomes important for setting standards and for scaling Agile adoption and Quality engineering itself.

When adopting QE in Agile, the validation activities are democratised within the squad members and extended team and based on the stage of the manufacturing process. This might not always be understood in tangible terms and in formal roles.  There is typically an expectation that everyone understands  quality and what is needed to build quality products, but that is not always true and often if there are lapses in Agile adoption or in the continued practice of Agile principles, quality falls to the wayside.  Everyone has the responsibility to contribute in the proving that the the work products built meet the requirements and standards. This might require reviews, tests, discussion  and feedback and drawing board actions. At each point a different lens of quality might be applied across the life-cycle from design to development through deployment and beyond. The Senior Quality Engineer helps establish quality responsibilities and ownership across the squads and the life-cycle. Further, she would expect that every role will adopt an automate first approach to drive rapid validation and ensure faster throughput. In this way, the combination of the teaming, the roles, and self-direction ensures that testing for quality does not become a speed breaker in the path to production. And the team looks to continually adopt new ways of building and engineering quality into every work product and deliverable.  

There are two interesting intersecting roles that are complemented by the Senior QE role. That of the Scrum leader (a.k.a. Team Coach)  and the product owner. As they drive the product road map, the release plan and the elaboration of the Epics and user stories, they are supported by the Senior QE to include the qualitative validation elements. The QE drives ownership and responsibility of quality into their very own functions as well.  While to some the  definitions of the ownership on quality might seem like the QE is vying for the place of the product owner or is hiving off the responsibility of quality to the product owner; that is not the case. The RACI below will clarify these doubts: 

\begin{figure}[h]
\includegraphics[scale=0.32]{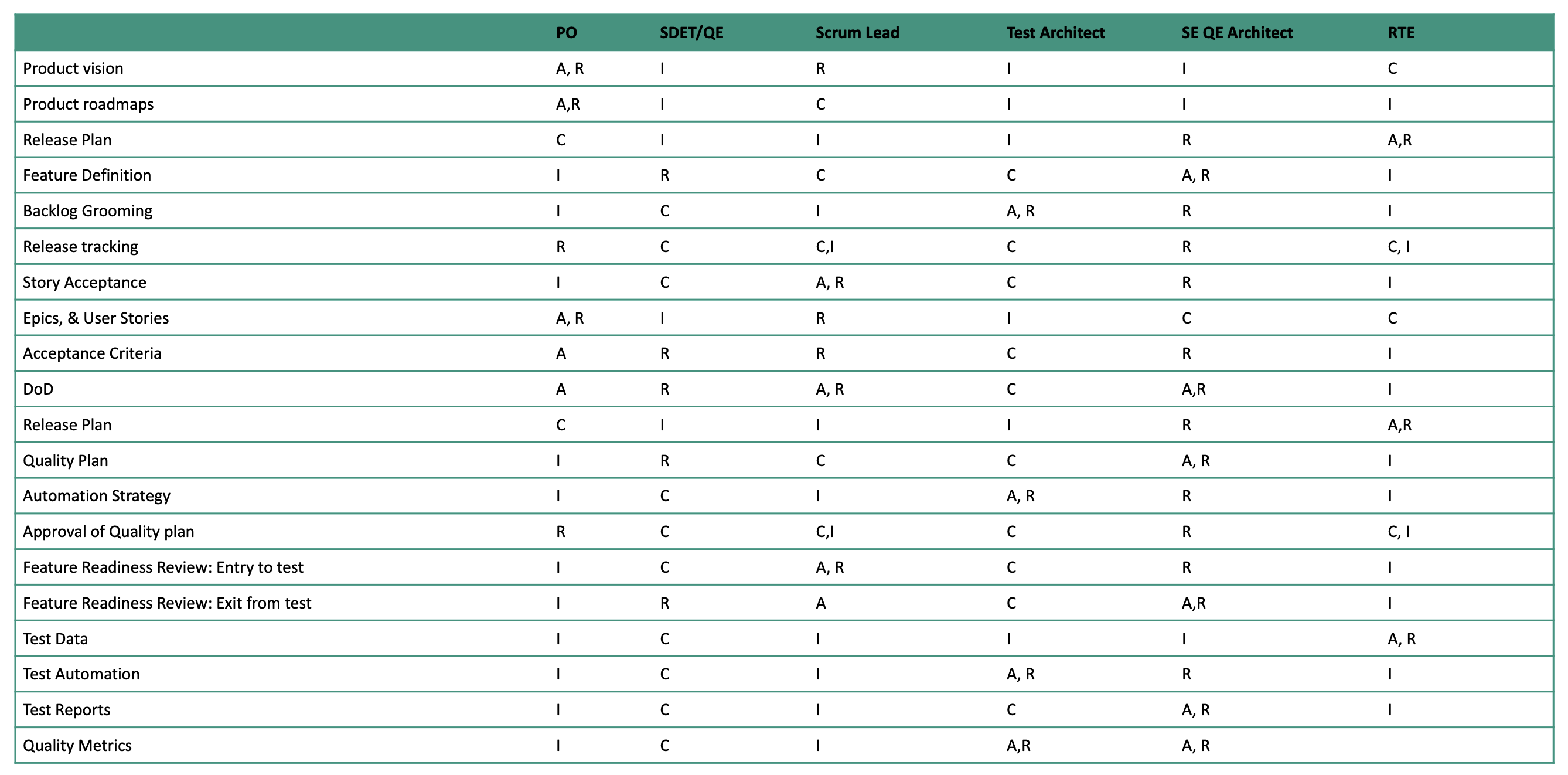}
\caption{Quality ownership across the squad}
\end{figure}

As you can see from the table there is a joint ownership on the quality elements as will be seen by how the epics and user stories are completed only on inclusion of the acceptance criteria and the definition of done. this inherently establishes the joint responsibility for the overall system quality. The Senior QE role is not a necessarily a permanent one and there is inherently a flexibility and fungibility in the perform role for driving QE. The aim is to enable quality ownership and drive the right mindsets, enabling adoption of best practices to build quality into the ways of working. This way quality engineering will be a strong enabler for speed and value instead of a check gate.

A question that seasoned testers and traditional software engineers will ask is, how dos this  new role of Quality  Engineering contract with that of the traditional tester who practices shift left? Also as we embrace the new model of quality being a shared responsibility how do we compare and contract with traditional roles of the project manager and beyond in delivering a quality deliverable.  

Let us look at the requirements and feature grooming process. The product owner typically brings in the larger perspective. The Quality Engineer wears the hat of the user or consumer itself and therefore has the unique role of representing the business value in a language that is not readable to the user but understandable by the system that is implementing it.   E.g. if we have a set of APIs being built to realise the business, then the business value of the API has to be translated into  API tests.  Those test might not be readable by a business user, but represents the needs of the business. Typically this is done by the QE or the developer and none of the other roles (e.g., project manager). While translating the requirements into system or API tests, the quality engineer will quantify and disambiguate the requirements themselves. this is done by creating tests for every end point, every interaction and detailing business actions and interactions as tests and detailing the inputs and boundaries of operation. This is the additional value that the Quality engineer will bring in. Typically a developer might write a similar test, but the tests would yield different results as the intent is different. the Quality Engineering is representing the user and hence develops tests to better validate the requirements. While the developer is utilising the tests to facilitate the construction of the solution. 

As the quality engineers disambiguate user stories or requirements in the process of writing the acceptance criteria and tests, they also inherently perform the activity of requirements prioritisation.  By this, we mean sequencing which requirements will be checked first, by ordering the tests.  This  mapping between the requirements and tests not only helps  surface potentially more complex defects earlier, it also helps prioritise the requirements.  Thus the quality engineer also feeds and augments the product owner and team coach (scrum leader) by helping identify the task that need to be performed and in defining the sequence in which the tasks need to be completed. 

As the quality engineer plays with the sequence of the tests, she also plays around with the sequence in which the requirements are to be built out and in the process she also influences the work assignments.  
The importance of this is really the cost implication, as these tests and the sequencing will result in deeper design deliberations and strengthen the build process.  It demands that the alternatives are considered upfront as a deliberate thoughtful collaborative  action on system build. This is cheaper than having to change the code or even worse the design later in the product build cycle. Oftentimes design deliberations are glossed over and ignored, as there is an implicit fear that there is no time or that the deadlines will be missed.  Analysis-paralysis does not set in when you adopt the model described above and the focus is on writing tests to define and prioritise the requirements.

The above approach driving deliberate design with feature prioritisation using tests can facilitate the adoption and scaling of domain driven design. This occurs as the definitions of tests refine and disambiguate the definition of the solution interfaces or APIs.  That improves the decomposition of the system with correct and clean definition of abstractions required for the Domain Driven Design approach \href{https://en.wikipedia.org/wiki/Domain-driven_design}{link} to succeed.

You would have noticed that in the process of requirements clarifications, disambiguation and sequencing, not only are acceptance tests written, but the system itself is build out as small and discernible lego blocks. The quality engineer, writes the tests, and is continuously establishing best ways to understand user needs, translate the user needs to system needs and into design definitions.   These functions  capture the movement of the classical tester into the quality engineering space and into role of interpreting the end user needs throughout the product life cycle. This elaboration will also cover non functional requirements and operational needs. Here the quality engineer brings in the construct of critical thinking upfront to help build the system from the requirements by breaking them down. 

Critical thinking is a structured and disciplined process of applying analytical skills to conceptualise, synthesize available material and augmenting understanding with inputs from prior experience, observation, reasoning and active inquiry / questioning. Critical thinking is a key aspect of constructing a solution that will actually work. It is sometimes neglected in the excitement of constructing the artifact.  It includes questions like what are the assumptions the solution is based on?  Are the assumptions correct?  Will the parts of the solution hold together to consistently provide the value the customer requires.  Introducing critical thinking when the requirements and design is being worked out reduces the risk of building something that does not work. The quality engineer plays a key role and is tasked with establishing right outcomes by wearing the hat of the critical thinking throughout the journey  of every user story that is defined, build and deployed. 

To summarise, the role of the classical tester moves to being more integral to the definition and development of the user stories themselves upfront. The quality engineer wears multiple hats from that of the consumer or end user and also of the critical thinker  as is required for interpreting the end user requirement throughout the product life cycle.




\chapter{Using QE to architect solutions that best represent the user}
\label{decomposition}




We discuss Agile quality related pitfalls and how to avoid and overcome them.  In order to do that we revisit the foundation principles of Agile development.  One such principle is to get the user in the loop of the development process. Typically, the user perspective is brought in via Design Thinking workshops where the user is present or is brought in as a persona around whom the software challenge is drafted and the solution is built out. Design Thinking is a model by which diverse participants come together to understand the problem holistically and work together to shape the solution. The process, if done well, should result in identifying new customers and new user stories at every level  and leads to an abstraction and a decomposition of the solution components.  

Getting the user in the loop of the development process  aims at  reducing risk. Note, we often use the words "user" , "customer", and "consumer"  interchangeably to mean the person or the human using the software, i.e., the external body or interface that is interacting with the software or code. The  definition of the customer or user changes to that of a consumer or producer of APIs, when we start translating the design to code. Applying design thinking principles would now mean developing the right empathy or deeper understanding of what needs to be done and how it will be contracted between the producers and consumers. 


When we consider the API personas of consumer or a producer, we start detailing the specifications of the APIs in terms of what is needed to complete a transaction and make it meaningful. We also establish the format and the trigger points of when the interaction will be conducted and what signals the completion. These details are important and a fundamental way of describing and publishing how your API is to be consumed.  

The API definition, establishes what you are going to get from the software but not how you are going to get it.  We thus build in the necessary abstraction or decomposition levels with the APIs.   This is an implementation of the hiding principle, i.e., the hiding of how the software is implemented from the consumer of the APIs.  In this way, the Design Thinking process, if well implemented, produces a solid, strong and well defined architecture for the software with the precise definition of APIs and user stories that use them, 

We now discuss the implementation of quality in the application in light of the hiding principle.  An application may include millions of lines of code which a human cannot grasp at once.  Instead, the hiding principal lets the human hide or ignore the details of an implementation through the introduction of interfaces or APIs. As a consequence, a typical execution of the application will result in calling dozens of APIs.  Now, these APIs may be developed by different developers and over different points of time. So an API may be produced by one developer and consumed by another developer and so on. This reiterates that the definition of a customer should be carefully considered and one should notice that it occurs dozens of times throughout the decomposition process. We revisit the concept of a user in the loop in the next section and the consequences to quality are analysed.  One may immediately notice that the Design Thinking workshop should not only be applied with end users that consume the external APIs of the solution. It also be conducted for the important internal APIs of the system that include producers and consumers that are not end users. These have indirect value that cannot be undermined or ignored. 


We also cover the following four Agile principles extracted from the \href{https://agilemanifesto.org/}{Agile manifesto } and how to implement them in a manner that ensures quality.  

\begin{enumerate}
    \item Individuals and interactions over processes and tools
    \item Working software over comprehensive documentation
    \item Customer collaboration over contract negotiation
    \item Responding to change over following a plan
\end{enumerate}

The book will take a stand that quality engineering is one of the important vehicles to the realization of the above Agile principles.  This will be elaborated in due course below.  For the time being notice that quality engineering produces artifacts that indeed enhance communication between individuals on working software and executable that brings everyone around a well-defined and concrete behaviour of the system and lets the system change with confidence using regressions.

\section{Who is a user?}

We develop software as a series of user stories that the working software will deliver. It is therefore imperative that we fist understand who the user is. We discuss the definition of a user as a consumer of user stories. We will choose a definition of a user that best implements the Agile principle of "user in the loop" and lets us reduce risk and increase quality. 

As we implement user stories we are developing code and the definition  of the  customer now moves to that of a consumer of interfaces.  The terms interfaces and APIs will be used interchangeably in what follows.  We reduce risk by precisely determining what the consumer/consuming interface relations are required to implement the user story.  Each consumer/producer interface relation defines a user, i.e., the consumer of the interface.  In addition, an interface  can have multiple consumers too.  A consumer-producer relationship is established by requests and responses.  To establish quality, we need to first understand the details of the requests and responses, i.e., the interface  details, and validate the functioning in the context of the user story being realised. This has to be done in iteration or recursively as each interface will also require additional interfaces to implement its part in the user story. 


\begin{example}
Consider a user story of reading a file by a file system.  The user story is implemented by searching for the file, opening the file and then reading its content.  This is implemented by three APIs, namely, a search API that locates the file, an opening API that checks for credentials, etc, and a read API that actually reads the file.  The Search File API is further broken into different search algorithms and APIs that depends on the where the file resides.  For example, search method for a local file is different than search method for a file that resides remotely.   We see that once the API is defined it can be further broken into additional APIs.   Therefore, the process of defining the APIs or micro services is iterative and requires that APIs are not not only consumed but are also able to consume other APIs.

We sometimes refer to a micro service as an \textbf{abstraction layer}.  The reason we do that is that we can focus on how the outcome is consumed, instead of on how the file is searched for in the above example.  The APIs that searches the file just finds it or not and reports the same.  We can be ignorant of whether the local or remote API were used to find the file.  Another term we apply is \textbf{a decomposition level}.  At a decomposition level the user story is broken into APIs that lets you achieve the business goal.  In our example, reading the file is decomposed into three APIs that search, open and read the file.  Finally, as any API can be further decomposed into additional APIs the \textbf{process becomes recursive}.
\end{example}

In a typical commercial application use case,  many interfaces will be invoked.  In addition, as discussed above, in order to implement the customer in the loop Agile principle, we will need to determine for each API call incurred by the use case, the producers and the consumers of that API, and have the identified consumer in the loop when the API is developed. 

To get a sense of how many APIs are used to implement a typical industrial application use case we consider the following back of the envelope calculation.  

Consider an application with a million lines of code.  Further assume that in a specific use case, 0.01 percent of the code, i.e., 10000 lines of code, are executed.  Assuming that each API is implement using 100 lines of code then 10 APIs are used to implement the use case.  Thus, at least 10 potential consumer producer relations needs to be identified and then customers defined and be "put in the loop" in order to follow the customer in the loop Agile paradigm.


\begin{example}

The following exercise further explains the notion of abstraction through interface definition.  Further, it enables humans to build software where details of the implementation could be ignored or "hidden".

\begin{exercise}
\label{avoidLosing:abstraction}
In order to implement the operation $x \times x$ a function, square(x), is implemented.  Two implementations are introduced.  
\begin{itemize}

\item One implementation is $return(x \times x)$.  
\item Another implementation checks if x is an integer or a float. 
   If x is an integer a loop is implemented that adds x to itself x times, namely $square = 0; for(i = 1; i \leq x; i++) square = square + x$. If x is a float $x \times x$ is returned.  

i.e., the code implements a type check, then calculates the value for either integer or for float and could also include error handling for invalid inputs.
   
\end{itemize}

Does it matter to the user which implementation is used assuming the implementation is correct?  How would testing the first implementation change compared to testing the second? 

See \ref{avoidLosing:abstractionSolution} for the solution.


\end{exercise}



\end{example}

We next examine the Agile manifesto in light of quality engineering.  Here is a paraphrase of the manifesto.   

\begin{itemize}
\item
Individuals and interactions over processes and tools.
\item
Working software over comprehensive documentation.
\item
Customer collaboration over contract negotiation.
\item
Responding to change over following a plan.
\end{itemize}

Interface bugs are one of the main sources of software bugs that have painful consequences to a system.  The other two categories are concurrency problems and memory leaks which are notoriously hard to debug.  The first Agile principle, namely "Individuals and interactions over processes and tools." emphasises the importance of communication between project stakeholders. Specifically, at each decomposition level of the software, the producer and consumer of the interface produced at that level may be different people.  Once they are different people, the chance of having miscommunication about the desired interface and as a result the chance of manifestation of an interface bug increases dramatically. When you translate this to your application design, the communication between the producer and consumer of an interface will define the interactions levels and relationship. The first Agile principle stresses that it is preferred that the consumer and producer of the interface at each level should directly communicate to ensure the quality of the interface.  It states that it is preferred over indirect communication imposed by process or a tool.



A word of caution is in order here.  Software engineering practices in general represents trade-offs between different options. The manifesto did not say "only direct communication" but suggested that we should "prefer" direct communication over tools and processes.  Tools and processes are useful and valuable.  Totally neglecting them as a consequence of attempting to implement the first principle to the extreme will probably lead to disastrous results.  Instead a trade-off that optimizes the processes should be found. This can be done especially whenever it is not possible or whenever it does not makes sense to have direct communication. This middle--way principle applies equally well to the other Agile principles. 

Implementing the first two Agile principles we thus want to identify the consumers and producers of interfaces at each abstraction level of the system and enhance their direct communication.  Note, in order to do that we need to define the abstraction levels, their interfaces, and identify the direction of communication i.e., identify the consumers and producers. More on how to do that in a way that enhances quality in \ref{strucutredModeling}.  We first examine how to enhance direct communication of consumers and producers of interfaces, assuming they were already identified.  The second Agile principle provides a guideline on one way of doing that - "Working software over comprehensive documentation.". Doing that effectively, will lead to overloading of the role of tests.  Tests are working software written in a formal language.  Communicating what the system needs to do using a test that consists of a series of interface calls makes the direct communication of consumer and producer of the users story implemented by that test precise. Thus, instead of writing tests only in order to find problems the role of tests change to also include the communication of what the system needs to provide to the consumers by the producers at each abstraction level.  The test being a precise and formal statement in a programming language that can be executed enhances communication, avoids misunderstanding and implements the principle of preferring working software over documentation.  We thus arrive at the Agile practice of test driven design (TDD) which we examine in more details in the next section. We will also examine the interplay between the TDD practice and the quality assurance practice of testing the systems in order to find bugs in the system.  We will see how the two practices overlap and enhance each other leading to engineering quality instead of only testing for it after the fact.  


\begin{exercise}
\label{avoidlosing:userstoryrefinement}
In your next sprint you are going to implement the plus operator between two numbers.  Initial design discussions suggest that you need to support the addition of any mixture of positive and negative numbers.   You proceed to design the following 4 tests.
\begin{itemize}
    \item 2+3
    \item -2+3
    \item 2+(-3)
    \item (-2)+(-3)
\end{itemize}

Next a review meeting is set and one of the review comments is that we also need to support  fractions and decimal numbers such as $2.3$.  There is an argument at the design level of whether or not we need to support imaginary numbers such as $2+3i$.  How would you update the test set?   Also which of the following resolutions of the imaginary number issue is in alignment with the Agile principles?

\begin{itemize}

\item Contact the user of the plus sign and find out if they will need imaginary numbers?
\item Implement with imaginary numbers to be on the safe side.
\item Implement without imaginary numbers.  If they are needed you will get a incident ticket and handle it then.  This way we will be able to develop faster.
\end{itemize}

See \ref{avoidLosing:userstoryrefinement:Solution} for the solution and explanation.

\end{exercise}


\begin{exercise}
\label{AvoidLosing:accessfilesystem}
Provide a few tests to a file system using the open, read, write and close API.  Attempt to emphasise what you would like the file system to provide through these tests.  Use the Linux man pages (e.g., \href{https://man7.org/linux/man-pages/man2/open.2.html}{Linux open man page}) to create a precise statement that could have been executed on a Unix implementation.  Could these tests be designed if the interfaces are defined regardless of whether or not a Unix implementation is available?

See \Ref{avoidLosing:accessfilesystem:Solution}
\end{exercise}



\begin{exercise}
\label{avoidLosing:AgileManifest}
You are giving a workshop on Agile and TDD (Test Driven Design).   The audience is extremely interested and would like to start implementing the new approaches in their development.  After the workshop they contact you with the following asks.

\begin{itemize}
    \item Please send us documentation on how to implement TDD.
    \item We have implemented tests that represents our next sprint.  Can you join the review and provide feedback on the TDD implementation.
    \item We wrote several tests that represents our next sprint and implemented the code.  Could you join our review and provide feedback on the TDD implementation?
    \item We have implemented TDD and are very happy, thank you!
\end{itemize}

Determine which of the requests above is consistent with the Agile manifesto.  Explain why.

\end{exercise}

See \ref{avoidLosing:AgileManifest:Solution} for the solution and explanation.


\section{Test first - testing and design}
\label{testFirst}

We first discuss Test Driven Design (TDD).  We highlight how a developer applies TDD in isolation to obtain efficient clean and high quality code quickly.  Building on the foundation of TDD we outline Acceptance Test Driven Design (ATDD).  In each abstraction level of the system, user stories are jointly defined by the consumer and producers of the interfaces at that abstraction level.  They are translated into test code before the user stories are coded.  They serve as an acceptance criteria for the completeness of the implementation of the user stories.  This approach creates direct interaction between the stakeholders that consume and produce the interfaces and prefers executing code, the tests, over processes and other less precise forms of communication.  Finally, we discuss how the tests produced by following the (A)TDD practices can be utilized to enhance testing done with the purpose of finding bugs. 

\section{Test Driven Design - the developer perspective}

We zoom in on the developer perspective of using tests for design (TDD).  Next section will consider the team level perspective (ATDD).


From the developer perspective, TDD is a practice in which tests are written before the code which fulfills the tests is implemented. The primary goal of TDD is generating a specification the implemented code should adhere to. By writing tests, the developer thinks about the requirements and design before implementing the actual code. A valuable side effect of TDD is that it results in significantly better code testing as compared to traditional techniques.   One way to intuitively grasp this is that typically as a result of the fact that code is written to fulfil the requirements of a defined test, probably every line of code implemented following the TDD process will be executed during the development process.  To sum up, in TDD, programmers work in small steps. First, creating a test, then adding its corresponding functional code.

It is instructive to consider the TDD process as a schema as follows.

\begin{verbatim}

T is the set of all tests previously defined
At the beginning of the development iteration we assume 
that all tests in T pass.

Begin development iteration
Write a test, t, to describe the requirements  
Run t to ensure that it fails
Write the code to make t pass  
Run t and all previous tests in T
if (t or any of the tests in T fail)
    return to the previous step and fix the code
else
    repeat the development iteration process

\end{verbatim}


We note the smell check of running the test that ensures the test fails before we write the code.  In addition to being a smell check that makes sure the test is strong enough - if the test cannot exhibit the fact that the code was not yet implemented it is worthless. Running the test before the code is implemented has the additional value of defining unimplemented interfaces precisely as the test may be calling interfaces that are not yet implemented.  This aspect will become central in the discussion of ATDD in the next section as our focus will shift from the individual developer to the developing team and the customer that consumes interfaces at a given abstraction level.   

Also note that we write code in order to fulfil the requirements of a test that fails.  We also run all previous tests developed for this abstraction level.  This implicitly ensures that our system is growing fulfilling more and more tests and their associated requirements.  It also implicitly assumes that our process is mature enough so that we can run tests automatically. Another point to observe is implicit but important. As we implement a test to fulfill the requirements of a test, and nothing else, we are attempting to create a minimal implementation of the requirements.  We thus attempt to avoid the creation of software that nobody requires.  We also reduce our development risk by incrementally developing the code in small chunks and always have code that passes the entire set of tests defined up to that point in the development process.    

But there is a risk associated with this seemingly perfect process. We may run into "local minima" - the fact that we are meeting more and more requirements does not mean that our code implementation is great.  In fact, the code will tend to become unnecessarily complex as we did not think about its implementation given the entire set of requirements up front.  This is dealt with by two additional important concepts associated with the TDD development process, namely, "bad smell" in code and refactoring of code which we discuss next.

\begin{exercise}
In order to write to a file we need to obtain its file name, open it for writing using the open interface and then if the open interface succeeds write to the file.   A code segment that open the file for writing includes 2-3 lines of code and we notice that it appears in ten different locations in our code.  Any suggestion on what should be done to improve the code?  Possible answers follow.
\begin{itemize}
    \item If the code is written in C use macros to execute the repeating statements, otherwise leave as is.
    \item Replace the repeated code with a function call that implements the sequence of opening the file and writing to it.
    \item Leave as is as creating an abstraction such as a function or a macro will impact performance.
\end{itemize}
\end{exercise}

\begin{exercise}
Our code accesses a shared variable using a lock mechanism.  It may look something like that $lock(m); x++; unlock(m)$.  We have a set of tests, T, that are supposed to check our application.  In order to determine if T is doing a good job, we substitute $lock(m)$ and $unlock(m)$ by empty functions that do nothing, i.e., $int lock(m){return(0);}$ and $int unlock(m){return(0);}$. We run the tests in T and they all pass.   What are the probable reasons for this?
\begin{itemize}
    \item The tests are running in a single thread and as a result are not testing concurrency. 
    \item There is no need for looking as no other thread is accessing the shared variable.
    \item Both cases above are possible.
\end{itemize}
Also, how is this questions related to the TDD methodology?
\begin{itemize}
    \item It is not related as concurrency should never be tested by the developer.
    \item $int lock(m){return(0);}$ and $int unlock(m){return(0);}$ are empty implementation of the lock interface.  If we are developing a locking mechanism the scenario above shows that the set of test T is not strong enough.  This matches the step in the TDD methodology if checking that the test fails.
\end{itemize}
\end{exercise}



\subsection{Bad smell in code and refactoring}

\begin{figure}[h!]
\centering
\includegraphics[scale=0.6]{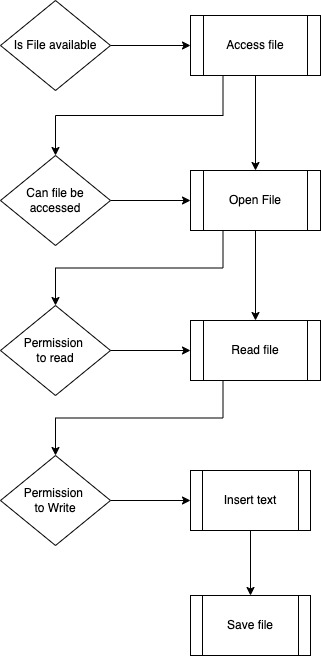}
\caption{Sequence of possible file access operations.}
\label{fig:fileAcccess}
\end{figure}

We briefly explain bad smell in code and how to remove them using refactoring. Bad smell in code is poor code that may be correct but it will be hard to maintain and evolve.  In other words, bad smells increase the chance that problems will be introduced in the code later on in its development cycle. To give a concrete example, consider a sequence of commands that repeatedly appear in the code.  For example, checking if the file is available, opening a file, checking that the file open operation was successful and if it  was, perform write and read operations on the file (see \ref{fig:fileAcccess}).  If a programmer, possibly new to the team, implements the access to a file in a new program location, we run the risk that the programmer will omit necessary checking, e.g., if the file was opened successfully or not.  Thus, the repeated occurrence of the file opening sequence of operations may be considered as bad smell in the code.  To overcome that, we can implement a function that executes the repeated sequence of file access operations including the check if the open was successful or not.  When the software functionality has to be enhanced and lets say, new code is to be implemented to perform a file write and read operation, we call the function that implements the needed sequence of operations.  Thus, the necessary checking, e.g., checking of whether or not the open operation was successful is done for us by the function and there is no chance that we miss it. 


The example above is an instance of the duplicate-code bad smell.  Essentially many bad smells in code has to do with a mismatch between the solving set of requirements introduced in each TDD development cycle and the design of the code, specifically the interface decomposition at each abstraction level.   This mismatch between requirements and design is introduced incrementally at each TDD iteration until it is hard to implement a new requirement expressed in a new test that fails.   At this stage it is natural to clean the code and ensure that the cleaned code still passes the entire set of tests developed thus far.   This process of cleaning is called refactoring.  We will next give some more examples of bad smell in code and discuss their associated refactoring.  Our description is in no way  comprehensive and the reader is encouraged to look up other  resources on code refactoring if they want to deep dive into the subject.  The examples below are meant to provide a better understanding of the concept of bad smells and refactoring and how they are related to the mismatch between requirements and design.   

\begin{itemize}
    \item Long method or functions implementations:  As a guideline functions implementation should not be too long.  One rule of thumb is that they should be easily read and understood within a few minutes.  A function that contains a thousand lines of code is hard to understand or further develop correctly.  Many a time such functions will include branches.  For example, in case of the Unix kernel implementation of searching for a file, the search starting point is dependent on whether or not the filename is an absolute file name, e.g., "/usr/bin/javac", or a relative file name "../JavaProgrammingResources".  The file search function may have 500 lines of code that handles the absolute file name input and 500 lines of code that handles the relative file name input.  This happened as the first few tests that were written in the TDD cycle had an absolute pathname as input and only then tests were introduced with the relative pathname input.  Once the code for the file search function is no longer easy to read,  e.g., a 1000 lines of code long, it is high time to stop and reflect on refactoring the file search function into two separate functions, one that handles the relative pathnames and another that handles the absolute pathnames.   Thus, the new decomposition and interfaces are made to better match  the requirements expressed in the tests written for the TDD cycle. 
    \item 
    Changes are made in too many places: Consider as an example, a procedural program where the data structure is composed of fields.  The data structure is initialized and logic associated with its fields is implemented in several locations in the code.  Whenever the data structure is changed by adding, removing or modifying the types of a field, all code paths that manipulate the data structure are to be inspected and the logic related to the new field added.  As a result, changes occur in many locations of the code.  For example, our software is focused on handling logic associated with traffic intersection in a city.  An intersection type is determined by the type of lanes that it intersects. Through a series of tests the programmer learns that there are the following type of lanes: car lanes, public transportation lanes, bicycles lanes, etc.  The programmer defines a LANE data structure and incrementally adds fields to LANE to handle each type of lane.  Each time fields are added to the data for a new lane type, logic is added in appropriate locations in the code to handle the new lane.  As a result the code is changed in several places in each iteration.  Recognizing this bad smell it is time to refactor the code, define a abstract data type that includes the LANE structure as its field and has methods that handle intersection logic given the intersection type. 
    \item 
    Use of nested IFs: Consider a scenario of determining the printing of insurance policy that includes applicable riders. This scenario might have a series of conditions based on state, age, policy type and so on. If you use a series of IFs to determine which policy riders are to be printed in the policy based on the client's characteristics and policy inclusions, you are likely to have very complex and intertwined and nested IF statements that will be hard to understand, maintain and extend as and when new riders and policy constructs come into play. Instead using a switch/case statement could help streamline and create an readable layout of the policy logic. The use of the default construct in the switch statement also enables the quick catch of a condition that the developer did not think about thus making the solution more robust. 


    \item 
    A function that has many input parameters, some of the input parameters:  For example, a hotel reservation system is implemented using a $reserve()$ method.  The reserve method grows incrementally, as before as a result of the order in which tests are introduced in the TDD development cycle, to a method with many parameters such as type of room, duration, time of check in, time of check out, with or without breakfast, type of payments, memberships of different clubs, etc.  Once the number of parameters makes the $reserve()$ method interface hard for comprehend , it should be recognized as bad smell and it is high time to refactor the code and break the reservation into different logical chunks that handle the room, food and payment separately. 

\end{itemize}

\begin{exercise}
Referring to the long method bad smell example above, specify
the two requirements that are mentioned in the file search operation. Having specified that, devise
the tests in the TDD cycles that will decrease the chances 
that the two requirements of the file search operation can be missed? Choose one of the
following answers.
\begin{enumerate}
\item One requirement is to find the file and the second is to find it only if you have permissions
to access it. The tests needs to cover relative and absolute filenames.
\item The requirement is to handle both relative and absolute pathnames. The test
should first focus on access permissions and only then handle pathname types.
\item The two requirements mentioned in the example are absolute and relative pathnames. It is better if tests with absolute and relative pathnames are both covered so that
the two requirements will be apparent from the beginning of the development cycle. 
\end{enumerate}

\end{exercise}





We also introduce the concept of a smell test.   A smell test in TDD, is a test that rapidly informs you that parts of the code has a bad smell.  This is akin to smelling eggs before you add them to your cake or pancake batter. Rotten eggs give off a bad smell, i.e., they stink. As soon as you break the egg, the smell of the egg will tell you if it is good or rotten, thus preventing you from adding a rotten egg to the batter and spoiling the dish.  So in software development, if you don't want badly written code components to spoil the final software you are building, you must recognise its smell using the smell tests. Don't let your final code stink! We must train ourselves to notice code that is not well structured right away when we see it.

Another metaphor that may help in the understanding of bad smells in code is found in the \href{https://en.wikipedia.org/wiki/Mikado_(game)}{Mikado} game.  Due to the way the sticks are arranged, it is hard to remove one of them without moving the others.  In code, if the code is not "well structured", it becomes hard to change it (move a stick), without breaking it (moving other sticks that you did not intend to move.).  When the code is not well structured we refer to it as a "bad smell" in code.  The idea behind the terminology is that when as in our example the same code sequence appears in many program locations, the fact that the code is not well structured becomes apparent.   More on that use of bad smell to improve quality in the section on reviews (see \ref{reviews}).

Conversely, think of building a tower of bricks as in the \href{https://en.wikipedia.org/wiki/Jenga}{Jenga game} to avoid "bad smells" and allow for a clear construct. Here you build a tower by pulling out and stacking up well-defined blocks of wood one on top of another. When doing that, the end-points, the recognition of weight-bearing bricks and stabilisers are apparent.  Your abstraction, e.g., micro services or API, should be the same, well-defined and structured, in a way that allows you to see dependencies and to rapidly call out bad-smells.



\section{Acceptance Test Driven Design (ATDD)}

With the understanding of TDD under our belt we are now ready to inspect the ATDD (Acceptance Test Driven Development) practice and appreciate how it implements the Agile principles while ensuring quality.  We use ATDD to express the requirements in a manner that makes it easier to understand the definition of done and enable accepting the implementation. Following the Agile principle of preferring individuals, interactions and customer collaboration over processes and tools, we  are using Design Thinking.  Design Thinking brings consumers, producers and other stakeholders together to work out the details of a given abstraction level resulting in the formulation of the appropriate APIs and the users stories.  Note again that the consumers of the APIs are not necessarily end users of the software solution.  

\begin{figure}[h!]
\centering
\includegraphics[scale=0.54]{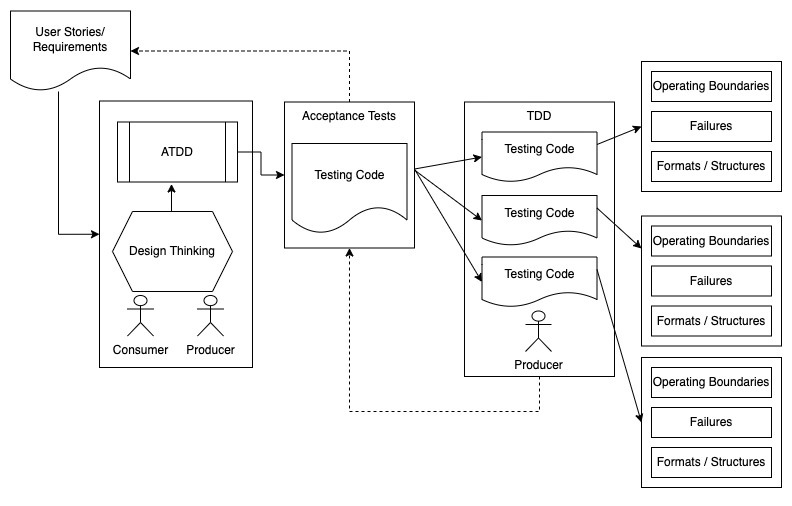}
\caption{From requirements to ATTD to TDD}
\label{ATTD_TDD}
\end{figure}

Next, to further implement the Agile principle of  preferring working software over comprehensive documentation, we formally define the APIs signature and write test code that implement the user stories that we want to realise at the abstraction level that is being discussed.  We are thus creating a set of tests that the consumers, producers and stakeholders agree will constitute an evidence of implementing the solution at the given abstraction level.  As in the case of the TDD process, we run the developed set of tests and ensure that they fail initially.  We refer to this set of tests as the set of acceptance tests, and hence the name Acceptance Test Driven Development (ATDD).  We are now ready to embark on the implementation of the design.  The developers use these tests as input to their development process and further break the requirements by introducing additional tests thus iterating through the previously discussed TTD process until all acceptance tests defined for this development iteration (sometimes referred to as a sprint) is completed.  The process is captured pictorially in \ref{ATTD_TDD}.


At this point one may ask the following question.  Traditional testing is done with the aim of finding problems in the system.  Above we have described the use of testing for driving the design, implementing the Agile principles and embedding quality into the development process.  Should we still do traditional testing?  What is the interplay between traditional testing done with the aim of breaking the software and testing done to drive high quality design and aid Agile delivery?  How do we avoid overlaps and redundancies?  All of this will be addressed in the next section.

Another natural question at this point is, how does TDD implement the Agile principle of 'respond to change' ?  The quick answer is that, automation of the acceptance test suite provides the confidence needed to make changes to the implementation of the abstraction level.  Once a change is introduced in response to a user feedback, the acceptance test as it is automated, can be easily run again to make sure that nothing in the software was broken.  The confidence in responding to change stemming from the automated acceptance test developed by following the ATDD process increases.   To underscore the point, if we did not have these automated tests that could be rerun rapidly to establish the feature developed we would be reluctant to include new features.  

\begin{exercise}
\label{avoidLosing:interfaceUseage}
Are all interfaces introduced in the software produced and consumed by different people?  Choose the most correct answer.
\begin{enumerate}
    \item Yes.  All interfaces are consumed and produced by different people
    \item No.  Some interfaces are produced and consumed by the same developers.
    \item No. Some interfaces are produced and consumed by the same developer.  Nevertheless, as programmers that maintain the code tend to change, eventually probably all interfaces of the solution will be used and further modified by more than one person.  
\end{enumerate}
See \ref{avoidLosing:interfaceUseageSolution} for the solution.
\end{exercise}

\begin{exercise}
\label{avoidLosing:testautomation}
Test automation means:
\begin{enumerate}
\item The test is well documented so that a person can easily follow it and enter the appropriate input to the software.
\item The test is a program. Once it is executed a person inspects the output to determine if the test succeeded or not.
\item The test is a program.  Once the test is run it produces an automatic output of whether it ran successfully or not.
\end{enumerate}
See \ref{avoidLosing:testautomationSolution} for the solution.
\end{exercise}






\section{Critical-Thinking-based testing}


One basic challenge of software development is that we introduce unintended behaviour as we develop it.   As a result, ATDD is not sufficient to ensure the correct behaviour of the solution and an additional activity of QE (quality engineering) needs to be brought into the process with the objective of breaking the system in order to ensure that the solution is correct.  We refer to that activity as critical testing.  The term critical testing is used as it is associated with critical thinking (see \href{https://skypeenglishclasses.com/teaching-children-english-using-the-six-thinking-hats-technique/}{thinking hats}).  The tests created during the agile sprint when following the ATDD practices can be utilized to facilitate the process of critical testing.  But first, we must go back to the fundamentals and examine why and how unintended behaviour is introduced when constructing the software solution.

\begin{exercise}
A code snippet below  includes 3 consecutive statements
\begin{enumerate}
\item
$if(x_1 > 0)~do~something;~else~do~something$
\item
$if(x_2 > 0)~do~something;~else~do~something$
\item 
$if(x_3 > 0)~do~something;~else~do~something$
\end{enumerate}

In how many ways can the software be executed?  Another way of asking the same question is how many different paths do we have in the code snippet?  Choose the correct answer below.
\begin{enumerate}
\item The code can be executed in 3 ways.
\item The code can be executed in 4 ways.
\item The code can be executed in 8 ways.
\end{enumerate}

Give inputs that executes the code in all possible ways.
\begin{enumerate}
\item It is impossible to answer the question as each variable has infinite number of possible values.
\item One possible value is that the three variables are positive,  and another possible  value is that the three variables are negative
\item Either input constitute of $x_1  = 1$ or $x_1 = -1$ and the same for $x_2$ and $x_3$
\end{enumerate}

Given that the number of lines in the code segment above is $n$ instead of 3 how many paths do you have through the code?

\begin{enumerate}
\item $n$
\item $n+1$
\item $2^n$
\end{enumerate}
    
\end{exercise}

In the exercise above, the code might have originally been meant to be executed in 3 different ways. But the combinations allowed as described above, can produce unintended results i.e., 8 ways of execution. The unintended behaviour is produced by combinations of variables or parameters we did not think about when implementing the solution.  Indeed, combinations of parameters that were not thought about when implementing the solution run a high probability of having an error.  The following exercise attempts to convey this intuition.

\begin{exercise}
Consider the following code snippet:

$read(x,~y);~if(x > 0)~x = 1;~y = -1;~if(y > 0)~x = x+y~else~y = x - y;~x = \frac{x}{y}$

which of the following  x and y input values will fail the system, name all correct answers.  
\begin{enumerate}
\item $x=1,~y=1$
\item $x=1,~y=-1$
\item $x=-1,~y=1$
\item $x=-1,~y=-1$
\end{enumerate}

In addition, what branch of the code segment above was not specified explicitly?

\end{exercise}

Another example of unintended behaviour follows.  
When opening a file for read, a file may be of different types. One of the possible types is a symbolic link that points to another file.  In that case, the read content should be parsed as a pathname.  If the implementation ignores this type of a file this will not be done.  
Also, as demonstrated in the above exercises, the number of such combinations may be huge (e.g., $2^n$ where $n$ is the number of binary conditions.). The situation seems hopeless but for the empirical evidence that errors are simple and typically have to do with a small set of parameter interactions.  This enables the application of Combinatorial Test Design (CTD) to the design of critical tests which we will discuss below. 

There is more than one way in which unintended combinations of parameters are introduced.  They could be introduced when an interface has several input parameters.  In addition, unintended combination of parameters can be introduced through environment variables.  For example, in the context of a file system, a process will have its current working directory from which the search for files start.   The current working directory can be of different types depending on the underlying file system type or if the directory is actually a mount point connecting two different file systems.  The mount point is another environment parameter that should be taken into consideration when conducting critical testing. Many a time, the fact that the directory could be of a different type might be ignored during the first implementation and could lead to errors in the implementation.  Another way to think about this is confound variables.  The type of the directory could be thought of as a confound variable, i.e., a variable that influences the correctness of the solution and might have been ignored in the original implementation.    

 You might have noted that following ATDD reduces the chance of misinterpreting requirements and miscommunication between stakeholders. While it also reduces the  chances of logical errors occurring, logical errors may still occur. To address this potential gap, we need a complementing approach such as critical testing.  We will discuss critical testing in more detail. Unintended parameters interaction is only one example of critical testing. 

Critical testing can be facilitated by using the tests introduced when following the ATDD process. Especially, as these tests make a specific choice of parameters. Varying the parameter values introduces additional tests that may be of interest and is a way to further check the solution.  One should focus on the variables on which the correctness of the solution is based, the rest of the parameters could be chosen randomly.  Even as we make a judgement that the rest of the parameters do not impact the correctness of the solution, we would still want to check the correctness of that assumption by lazily covering them with randomly choosing their value each time we run a test.   This approach makes our critical testing process stronger. We further clarify the approach in the following exercise.

\begin{exercise}
\label{CriticalTesting:ATTDReadFile}
    
Assume our ATDD test opens a file for read, reads and then closes the file.   Specifically, here is the test code snippet.
\begin{verbatim}
fd = open("write", "../myFile");
if(fd == 0) //open for write succeed
    write(fd, "it is my file");
fd = open("read", "../myFile");
if(fd == 0) //open for read succeed
    buf = read(fd);
if(buf == "it is my file")
    print("success")
else
    print("failure")
\end{verbatim}

What other tests would you introduce to check the system by varying parameters of the above test?  Choose all correct answers.

\begin{enumerate}
\item I will change the size of the file.
\item I will change the size of the file and the types of the filename (also pathname of the form "/myfile")
\item I will change the type of the file.
\item I will change the type of the current working directory of the process.
\item I will explicitly close the file or not.
\end{enumerate}

See answer in \ref{CriticalTesting:ATTDReadFile:Solution}

\end{exercise}

Combinatorial Test Design is one efficient method to design critical tests that cover all possible variations and paths and produce the minimal number of tests required to fully test the software. This technique will enable you to be more more effective and efficient as it will help you avoid duplicate tests while enabling coverage of all parameters in the most effective manner. You might use other techniques as well, such as defining customer journeys, user provided test inputs, fish bone analysis, what-if analysis, 5Whys, inclusion of boundary value conditions,  equivalence partitioning, etc. These methods will help you understand edge cases and variations in test paths, but might not always provide the 100$\%$ coverage and assurance that the tests created using CTD does.   For further information on CTD see the following short methodology white paper: \href{https://www.overleaf.com/read/hrcbrhrzwnfq}{link}. 

As critical testing is closely related to critical thinking we next highlight some best practices to help you build basic critical thinking capability and skills.  Critical thinking is all about flipping your mind set.   Instead of being positive and attempting to construct a solution you want now to attempt to become the devil's advocate and if possible break the solution in a structured manner.  That mind set when adapted effectively will lead to a strong set of critical tests.  Fundamentally, critical thinking has to do with examining the solution and its assumptions.  The following questions are highly useful to follow as a checklist. 

\begin{enumerate}
\item Does the solution make sense?
\item What are the assumptions under which the solution should hold?  
\item Will the assumptions hold?
\item What is vague or unclear about the solution?
\item Are there any implicit assumptions that can be made explicit and checked?
\item Any hidden parameters that can impact the correctness of the solution?
\item What are all the points of variations of parameters and states?  Which one of them are thought not to impact the solution?   Does it make sense and can the assumption be tested?
\end{enumerate}

Another area to address with critical thinking is understanding and addressing bias in the requirements and implementation. Biases are standard and default ways in which a system or requirement is understood and implemented. Critical thinking helps build tests that break system boundaries by questioning biases based on which they are built.  This helps us build  "Negative Tests" - that is tests that look to address what the system should not do. Negative tests are good for trapping issues with error handling, memory leaks and missed use cases and prevents misinterpretation of user stories.

Questions that will help you develop tests that address biases in the system include: 

\begin{enumerate}

\item what are the inherent biases in the system?
\item what can flip the system to work outside the bias boundaries? 
\item what are a possible tangential parameters or variables that can flip the system? 
\item what will make a sequence of actions illogical? 
\item what are good alternatives or synonyms to include in the tests?
\item what can create a basis for more data to test the system paths?
\item what are the negative tests that will likely break the system?
\item how can the system be misinterpreted?
\item what will result in a dead-lock situation in the functional implementation? 
\item can I use the system concurrently and will it lead to race conditions?  
\item what are the possible single points of failure?
\item what is the hierarchy of functional calls, that I should break?
\item What are the alternatives? Trade-offs? 
\item How can I drive unintended consequences? 
\item What could be the possible list of unintended consequences? 
\item How can the system be driven to exhaust its resources?
\end{enumerate}

Use these above questions in iteration for each user story and user story implementation and over time you will find that you are able to critically evaluate the system needs upfront and thereby will be able to build negative tests that can even help define the system better upfront.

A challenge for your agile scrum leader or team coach will be use these the tests and outcomes of your critical thinking process to build out the needed details on the user stories, and in the backlog management. 


\section{Design thinking as an internal abstraction level}

Notice that we emphasize the use of design techniques at any meaningful abstraction level of the system even if the abstraction level involves only internal consumers and producers.  In this section we will highlight how and when this is done and the differences that apply compared to a design thinking workshop done with the end user of the solution. 

Design thinking is about understanding how to build out a solution by first understanding the need for a solution from the perspective of the end user. Empathy and process maps help appreciate the rationale of the requirements and thereby enable better envisioning of the expected outcomes of the solution. This in turn helps in crafting of detailed user stories.

\section{The oldest trick in the book – reviews!} 
\label{reviews}

A review is typically a formal assessment or appraisal of a work product, with the intent to validate and institute change if necessary. Reviews need to be considered as a validation activity as well as a constructive critiquing phase. In software development, reviews are early ways to gain insights of the quality being built. Reviews apply to every work product in the life cycle. It is also a  first step to verify correctness of the communication across stakeholders. It needs to be ubiquitous across all life cycle activities and continually be the voice-in-the-head providing course correction and assurance.  Reviews thus become pivotal to driving a faster path to production and need to be integrated into the DevSecOps pipeline processes.  Typically, reviews tend to be static validation in nature and not a  runtime process. Thereby, they are less expensive as they provide early quality insights without running the code and therefore you do not need a test environment to realize them. Listed below are some of the different reviews that need to be included.

\begin{itemize}
    \item User story refinement 
    \item Build consensus on the definition of done (DoD)
    \item Review of non functional requirements (NFR) 
    \item Code review (static code scans)
    \item Security scanning (Static Application Security Testing (SAST))
    \item Build consensus on the Definition of ready
\end{itemize}

Checklist-based reviews, compile-time reviews and inclusion of design patterns, adherence to coding standards and other best practices are  the validation aspects of reviews. There are many tools in the market that provide code reviews and code quality assurance as a static testing activity in the pipeline.

Reviews are also implemented by pair programming and by the use of Gen AI based coding and code generation solutions. 
Explore how generative AI solutions can help you write better structured code snippets.

\section{Defining and validating abstraction levels}
\label{strucutredModeling}

Recapping our discussion thus far, we recursively, at a given abstraction level, define the use cases and capture the interfaces that will meet the user's needs at that abstraction level.  To do that we follow the ATDD process implementing the Agile principle of preferring working software over comprehensive documentation.  We also prefer that individuals will interact over processes and tools as we bring them together to discuss the use cases and interfaces utilizing the design thinking instrument.  As a result, we are able to respond quickly and with confidence to changes as the ATDD naturally results in an automated regression suite at the end of the implementation sprint and enables quick response to changes when needed.  


\begin{exercise}
Why does having an automatic regression help respond to changes?  Choose the most correct option below.
\begin{enumerate}
    \item Regression defines how the system should behave and thus provides a reference point.  Updating the regression helps define the next sprint.
    \item Regression tests let you quickly check if nothing is broken due to the software update thus raising the confidence that you will be able to make the right changes.
    \item Both answers above are correct. 
\end{enumerate}

\end{exercise}

One of the key principles of software engineering is that there is no silver bullet practice or method, instead based on context and need, we develop fit-for-purpose practices. This includes making trade-offs in order to reach the best possible practice.  This principle is next applied. 
To better integrate quality engineering into the Agile process we make a trade-off and choose not not abandon the documentation process all together.  Reexamining the  
\href{https://agilemanifesto.org/}{Agile Manifesto} first two principles:

\begin{enumerate}
    \item Individuals and interactions over processes and tools
    \item Working software over comprehensive documentation
\end{enumerate}

A common myth of Agile methodology is that there is no documentation.  Re-read the above two manifests as we elaborate how lightweight and fit for purpose processes and documentation will enable successful adherence and thereby quality manufacturing.

One notices that our choice of not abandoning the documentation process, does not contradict the manifesto. While, indeed the manifesto suggests to prefer working software over comprehensive documentation, it does not means that documentation is excluded altogether.   In fact, documentation can and should facilitate the realization of the first principle, namely, the preferences of individual's interaction. As our focus is quality and correctness, we introduce a set of light weight artifacts that help facilitate the definition and validation of the requirements and their design through communication between appropriate stakeholders. A few examples are the inclusion of design thinking, use of murals and mind-maps to document user journeys, context diagrams,  open API specifications, and such to promote interaction between individuals to establish a valid understanding of the requirements, design elements,  points of interaction and integration and dependencies.  Our style of further facilitating individual interaction is by using the oldest trick in the book of software engineering, namely reviews.

Our approach includes elements which we elaborate on in the following sections.  

\begin{itemize}
\item A clear distinction  between what the solution needs to do which we refer to as "requirements" and how it is done, which we refer to as "design".
\item Awareness of the pitfalls involved in defining requirements.
\item Context diagram.   Clearly define the abstraction level and its relation to the rest of the solution.
\item Interface definition.  The use of design by contracts. 
\item Use cases. Determine how use cases utilize the interfaces to satisfy the needs of the users.  Note that this stage and the previous one leads to the definition of the ATDD test suite for the sprint.
\item The application of Combinatorial Test Design (CTD) to the use cases and the context diagram.  Note that this stage feeds into the Critical Thinking activity. 
\end{itemize}

We finally discuss a particular review style that brings the stakeholders together to interact and determine  if the above artifacts are consistent and whether implementing them will meet the users needs.

Note: In Agile terminology, requirements are typically referred to as epics, features and user stories. This is to translate requirements into bite-sized chunks of what the solution needs to do i.e. the needs of the customers that are amenable to  being distributed across sprints and squads to own and drive. We will use these terms interchangeably or simply we refer to them as requirements.  


\subsection{Distinguish between what the solution needs to do and how it is going to be done}

We now take a deeper look at an abstraction level.  When attempting to define an abstraction level a broad distinction arises between two types of activities. Whenever “what should be done” is the main focus of the discussion one should focus on the definition of requirements, whereas when the discussion is focused on "how it should be done” design artifacts such as the context diagram, use cases, interfaces and their contracts and CTD should drive the discussion.   Failing to distinguish between these two type of activities and failing to follow best practices for requirements elaboration and design definition, increases the chance that we will build a solution that either the user does not need,  or a design that is not maintainable. both of which are in fact  typical risks in the software development process.

We start with some basics about requirements definition.  Once we have identified the customer at a given abstraction level which could be the end customer or an internal consumer, the basic question that we should address is does a requirement reflects the customer’s need?  Naturally, and following the Agile principle of putting the customer in the loop, the customer should be part of the definition process of the requirements for the given abstraction level.  Acknowledging that the customer may be unclear on its needs upfront, putting the customer in the loop of the requirement definition, while having a huge benefit, also results in some potential risks which needs to be managed as we will see below(see item \ref{customer} in the below list.). 

\begin{exercise}
This exercise requires background in computer science and compilers.  There are two very broad categories of programming languages namely, \href{https://en.wikipedia.org/wiki/Imperative_programming}{imperative and declarative languages.}   In addition, a given programming language needs to be assigned semantics that defines its meaning.  Again there are two broad categories of semantics, \href{https://en.wikipedia.org/wiki/Semantics_(computer_science)#:~:text=In%20programming%20language%20theory%2C%20semantics,program%20in%20that%20specific%20language.}{declarative and imperative.}  Discuss how the declarative and imperative languages and their semantic is related to the notion of requirements and design defined above. 
\end{exercise}

Typical pitfalls to be avoided  when attempting to define the requirements of an abstraction level are listed below  along with  some examples and tips on how to overcome them.      
\begin{enumerate}
\item \label{customer} A customer may be describing its requirement in an indirect manner that confuses the real requirement that the customer has.   For example, the customer may say that we need to enable multiple copies of servers on different nodes.  Further investigation indicates that the customer requires that the system will continue to operate in the face of one of the nodes failing.   This can be handled in multiple ways and not necessarily though the introduction of another node.  One way to handle this type of an issue is to work with the customer to reach an agreed definition of the requirements that declare  what is needed.  In our example, uninterrupted operation of the solution in face of failure of the node would be such a declarative statement of the requirement.  It avoids specifying how the requirement is realized.  It let us explore different design alternatives that meet the requirement and make an optimal design choice.  We can summaries this best practice as follows "use declarative and not imperative language when defining requirements".
\item Another typical issue with requirement definition is that it is not clear who is the originator of the requirement.  Remember that we only want the solution to be as complex as needed.  We do not want to implement features that nobody will use.   A good guideline here is to understand who will be the first user of the new use case we are introducing.  
\item  Requirements must not be open to interpretation and they must be measurable. Precise words and words that have only one meaning should be used in their definition as well as exact numbers. One should avoid terms such as  “approximate”, “maybe”, “about”, etc and prefer the term “must” instead of “should”.  To summaries, one should Avoid vague and general terms such as “good performance”.  For example, instead of saying that a server needs to be reliable one should say something like the sever is allowed to go down once every three years.  Instead of saying that an editor should be easy to use one could say that the editor should follow a what you see is what you get policy. 
\item Will you be able to demonstrate that the requirements are satisfied?  Are the requirements testable?  Requirements that cannot be tested through our ATDD process are not requirements and should be discarded or modified to be testable.  
\item Separation of concerns is a best practice in software engineering.  Defining the requirements is not a project management activity; it does not and should not include prioritization of requirements.  That should be kept as a separate activity.  Words included in the requirements definitions that hint on prioritization should be avoided.
\end{enumerate}

\begin{exercise}
Based on the proceeding discussion in this chapter define the following terms:

\begin{enumerate}
    \item Design
    \item requirements
    \item Customer
    \item User
    \item Abstraction level
    \item end user
    \item internal user
    \item Consumer
    \item producer
\end{enumerate}

Search the internet for definition of the terms design, requirements, customer, user.  Compare with the definition used in this book and discuss the major differences and similarities. 

\end{exercise}

The last exercise teaches that the terminology acquires specific meaning when a new context is defined.  An example of that is the term thread.  A thread is a cord, but in computer science it is a process that share memory with other threads!  One needs to be aware of the terms being used when describing the requirement and design and their specific meaning.  We deep dive into this topic in the next section.
\subsection{Using, defining and reviewing terms}

In this section we discuss the definition of new terms during the requirements and design process.  New terms definitions are kept in a glossary. As was clarified in the previous section defining new terms serves to disambiguate and make precise the meaning of terms used in the context of the abstraction level being defined.  As we will shortly see the definition of a glossary of words serves two additional purposes.   It helps avoid terms that are ambiguous or do not belong to the current abstraction level but to its implementation.  For example, at the current abstraction level we may be discussing a number variable.  At this level whether the number is implemented as a byte or a float is not relevant.  In fact it is an implementation detail that should be avoided as the decision of whether it is a byte or a float should be made later on when implementing the interfaces defined in this abstraction level.  We thus also use the term "negative glossary" to mean terms that should not be used when defining the current abstraction level.

We start with an anecdote. One of the authors of this book gave a presentation to the team on CSP.  Many questions were asked but they did not make sense and seemed unrelated to the author.  After a while, it became clear to the author that the team had the term Constrain Satisfaction Problem (CSP) in mind due to their background in formal verification of hardware while the author had the term Communication Sequential Processes (CSP) in mind.  This anecdote highlights how undefined terminology and especially overloaded acronyms can cause huge communication barriers.

\begin{exercise}
Search CSP on the internet in various ways and list all the possible meaning you were able to extract for the acronym.  
\end{exercise}

Thus, when defining requirements and design it is important to define a glossary that captures the terms that have specific meaning in the abstraction level being defined.  Doing this helps ensure a common understanding of the abstraction level between the stakeholders. Introduced acronyms is an easy give away and must be defined but they are not the only new terms introduced in the requirements and design process.  For example, an operating system design will probably include terms with specific meaning such as process and memory that needs to be defined.   In the context of an operating system design memory does not mean the human memory and process does not mean a series of steps that are used by an organization!

\begin{exercise}
What is the meaning of a process and memory in the context of the design of an operating system?  Search the internet if needed to obtain correct definitions.
\end{exercise}

Good glossary definitions are short precise and consistent.  To be effective they focus on the new terminology introduced for the current abstraction level.  Thus, a good glossary is small. In addition, while creating the glossary one should avoid assuming that terms are known and when in doubt add a term to avoid the risk of misunderstanding. Global glossary that apply to the entire solution may be applied and reused.  Awareness to the standard use of terms and the use of terms within the company should be applied to avoid confusion. 

We next examine the concept of a negative glossary.  A negative glossary highlights terms that should not be used for describing the current abstraction level.  It is composed from generic words that should never be used as they cause confusion and specific terminology that does not belong to this abstraction level but to the implementation.  Thus, the use of a negative glossary helps better define the abstraction level. The negative glossary definition is not a static dictionary of terms; but is also a powerful review mechanism, that gets built into the communications and interactions that help define the requirements or user stories better.  It helps set a benchmark of what good looks like. 

Tip: Set aside 3 to 5 minutes in daily scrum calls, to review and update the negative glossary for your project. Based on a reading of the user stories, actively look to adding terms to the negative glossary and removing those terms from the documentation of your Epics, Features and User stories. You will find that the quality of understanding the requirements improves and thereby of its quality and detailing of the requirements improves.

\begin{example}
We next consider examples of negative words.  We list an example sentence and explain why the word should be avoided when defining requirements.
\begin{enumerate}
\item "We want to run a backup on Sunday, Tuesday, Thursday, etc."  The negative word is "etc".  It indicates that we did not specify the rule and left the requirement ambiguous.  A possible rule here might have been "We want to run a backup on alternate days of the week.".
\item "The response time of the system should be adequate."  The negative word here is adequate.  In fact, we are not quantifying what adequate is, thereby leaving it ambiguous.  Note that another indication to the weakness of this statement is that we can not write a test that will check if the system's response time is "adequate".
\item "The system needs to provide tokens to user A and/or user B".  The negative term is and/or.  The conditions under which only A receives the token, only B receives the token, and both of them receive the token are not specified. 
\item "The solution should approve the withdrawal of money if it is appropriate to do so.".   Here "appropriate" is the negative word.  In fact, it is not clear by this statement under which conditions it is appropriate to approve the withdrawal of the money. 
\end{enumerate}
\end{example}

  Here are some  examples of  generic negative words that should not be used in practice, adequate, and/or, all, as a minimum, as applicable, as appropriate, be able to, be capable, between, but not limited to, capability of, easy, effective, efficient, etc, every, flexible, if practical, improved, including, intuitive, large, maximize, minimize, normal, not limited to, Often and Various.  The list is not complete, is meant to provide a starting point, and its customization to the need of a specific project should be considered. One caution is that negative words are not universal and sometimes their use is OK, provided the context is set.

We now turn our attention to terms that should not be used to describe an abstraction level as it belongs to the implementation.  Avoiding such terms is extremely important.  It helps reaching a correct declarative level of description, implement the basic software engineering principle of hiding and avoid making design trade-off too early and unintentionally thus creating a sub optimal solution.  The following example clarifies the concept.

\begin{example}
In general classical abstract data types such as lists, trees, stacks can serve as a good example.  To be specific consider the list abstract data type.  A list is a sequence of records. Its typical operations include the creation of a list, adding and deleting an element from a list and traversing each of the list's records.  The list is abstract in the sense that we are not interested in how we are implementing the list.  One may implement a list using consecutive memory allocation (an array) or non-consecutive memory allocation (pointer implementation).  When describing the list abstract data type one should avoid the description of the list implementation.  Thus, words like array and pointers are negative at the abstraction level that describe the list and its operations. 
\end{example}
\

\begin{exercise}
Provide detail definitions of the list interfaces, namely, create, add, delete, and traverse avoiding the use of terminology that has to do with the implementation of the list.
\end{exercise}

\subsection{Requirements elaboration}

How do we go from requirements of the entire solution to smaller chunks that are ready for design elaboration and implementation?  User stories are applied to do this.   As we are elaborating requirements so declarative language is still in order.   Specifically, we apply the following style of user story definition: As a $<$type of user$>$, I want $<$some goal$>$ so that $<$some reason$>$.

\begin{exercise}
You are given the following user story: As a system programmer I want to store and retrieve my data from a file in persistent storage so that I can handle the users of a database.  Which of the following options are elaboration of the given user story?
\begin{enumerate}
\item As a system programmer I want to add new users to my database user's file including their names and credentials so that I can determine their allowed scope of operations.
\item As a system programmer I want to remove users from database user's file so that I can disable access of users to the database if it is no longer allowed.
\item Both options above are elaborations of the given user story.
\end{enumerate}

\end{exercise}

Using the user story format enables the break down of requirements into more visual and perceivable definitions of what the system or software should do.  We use the following terminology to describe this elaboration process of the user stories.     
The terms Epics, Features and user stories are used to establish the needed elaboration, context and detail.  These terms inherently establish a hierarchy and an abstraction to a context. The Epic is an impressive or large body of work that illustrates the system's functional capabilities required to meet the client needs.   It is the overarching capability that the system or functionality needs to achieve. It is typically understood that the epic will need more than one sprint cycle and multiple squads for realisation. It gives the broad sense and direction of what the end user wants.   The epic may include one or more all encompassing user stories.

An epic being a large body of work is typically composed of stories that are contextualised into an achievable piece of work.

\begin{exercise}

Write down the epics and user stories for a mobile application that will create a bridge between university alumni and current students of their Alma mater aimed to enable interactions to help students.
\end{exercise}

It is important to emphasize that as we drill down from epics to more and more detailed user stories, that the artifacts we produce are still declarative within their context.
Next, the work or activities they need to perform are elaborated as actions to be taken to help realise the work products as defined in the user stories.  At this level we turn our user stories from declarative to imperative.  We do this by translating the detailed user stories into a series of API calls effectively creating an abstraction level that id defined by a series of use cases.  This step transition our declarative description of the solution to an imperative one. 
 More details on user cases in the design section. 



As we go through the elaboration process of more and more refined user stories until enough details are available to define use cases the following pitfall arises. Project management and work assignment is confused with the requirement and design process. As a result task allocation and requirements break down and user stories are mixed. This results in a divergence of focus from capturing the clients needs to task and work assignment mind set may lead to degradation of quality and overrun on budget as you are tracking and focusing on the wrong thing.  


A common pitfall is to mix up work break down and task allocation and requirements break down and user stories. This results in a divergent focus on what needs to be done versus what needs to be accomplished by the system.

\begin{exercise}
We are given the following user story: a user in an online bookshop orders a book.   This is further elaborated through our development process into the following items.
\begin{itemize}
\item A development activity that implements the search for a book.
\item An activity that implements the checking that the book is in the inventory
\item A development activity that implements the credit validation of the customer
\item A design element that defines three APIs.  One that search the book, another that check if the book is in the inventory and another that  validates the credit of the use
\item A test that lets a customer buy a book in the online bookshop
\item A test that checks if the book is in the inventory
\item A test that checks if the customer the credit
\end{itemize}
Which of the above items can be part of an acceptance criteria of the solution.
\begin{enumerate}
\item All of the items above are part of the acceptance criteria.
\item Only the end to end test described above is part of the acceptance criteria
\item Only the end to end test described above is par of the acceptance criteria but the tests for the individual APIs can also be added to the acceptance criteria
\item The proceeding items are correct.  
\end{enumerate}
\end{exercise}

Note that the lesson of the above exercise is that development activities are not acceptance criteria.  In fact for a user user perspective only a user case that is realized in a test can serve as the base for an acceptance criteria.  Other secondary acceptance criteria may apply such as non function performance indicators and API documentation that helps consume the desired use case.

\subsection{Design - focusing on the how in the abstraction level being defined}

Up to this point in the exposition, we have focused on requirements, i.e., the definition of what the abstraction level should do.  We next go from the declarative to the imperative and start defining how the requirements are met. In other words we start defining the design of the system.  Typically this is done through use cases implemented by a set of interfaces that hide implementation details at the current abstraction level.

Good design makes trade offs with respect to conflicting objectives but a key objective of the design process is to notice and analyze these trade-offs. For example, one such trade-off has to do with performance.  The following exercise provides an example.

\begin{exercise}
A solution was developed under the assumption that writing to the local disk is much faster than writing to a disk on other machines that are located in the organization local network.  Under these assumption the solution had a cache that kept frequently accessed data on the local disk.  Although the cache introduced complexity to the solution and some windows of potential inconsistency and data loss it was accepted as a reasonable trade-off needed to meet the required performance requirements of the solution.  
After a recent upgrade of the local network measurements are now showing that writing to disks of computers in the local network now has comparable performance or sometimes even faster than  writing to the local disk. What kind of new trade-off can be realized to improve the solution?
\begin{enumerate}
\item Remove the cache, it is no longer needed.
\item Create a mechanism that measures if local or remote disk access is faster and make a decision accordingly.
\item The two options above could be explored to simplify the solution.
\item Nothing should be done as the software it working and you never refactor software that is working.
\end{enumerate}
\end{exercise}

Moving on, in addition to identifying the trade-offs, one should try and understand if the abstraction level is well chosen.  The suggested abstraction level decomposes the system into interfaces that implements the use cases.  One should inspect whether the interfaces are loosely coupled which is the hallmark of a good interface choice. Loosely coupled here means that data that belonged to the implementation of one interface is not repeatedly sought for by another interface.  Another criterion is complexity.  We want the decomposition level to be as simple as possible. We thus do not want it to implement things that are not required by the use cases.  One last effective sanity check of the abstraction level; check that you can understand the abstraction level without considering other level, especially implementation details that belong to the lower level of abstraction.


\begin{exercise}
Consider as solution that solves a quadratic equation $ax^2+bx+c=0$.  The solution includes the following interfaces.
\begin{itemize}
\item boolean solutionExists(int a, int b, int c) - determines if a solution exists or not.
\item findSolution(int a, int b, int c) - this interface is invoked only if there is a solution.  
\item findSOlutionLinear(inb b, int c) - this interface is invoked only if there is a solution and $a = 0$.
\item findSolutionQuadratic(int a, int b, int c) - if $a \ne 0$ then and there is a solution then this interface is invoked.
\end{itemize}

How would you define the interfaces to avoid the need of calling findSOlutionLinear() and findSolutionQuadratic() from findSolution()?  Chose best option below.

\begin{enumerate}
\item Once solutionExists() determines that a solution exists, the solution will $a$ is 0 or not and invoke findSOlutionLinear() or findSolutionQuadratic() accordingly. 
\item Determine if $a = 0$, invoke findSOlutionLinear() or findSolutionQuadratic() accordingly.  Let the two interfaces handle wether or not the solution exisits.
\item There is no way to avoid the two calls.
\end{enumerate}
 
 Discuss the trade-offs between different ways of defining the interfaces in light of the principles discussed in the previous paragraph.

\end{exercise}

Many times it is best to begin the design process with an activity named "an overview".   An overview should briefly capture what the abstraction level should do (requirements) and how it is done (design), i.e., the interfaces that will decompose the current abstraction level and the use cases that use these interfaces to implement the requirements.  In brief the customer and developers of the abstraction level should be invited to participate in the overview activity in addition to any other impacted stakeholders. More details on the choice of stakeholders in the review section \ref{reviews}.  
Overview is the first step in selecting what design artifacts to document    In addition, the abstraction level overview should defined the boundaries of the solution which we discuss in more details in the next section.  Latter on the  overview activity will be repeated when the abstraction level is reviewed to help stakeholders come up to quickly understand the abstraction level and be effective in reviewing its design. 

\subsection{Context diagrams}

A Context Diagram is the highest level view of an abstraction level. It demonstrates the abstraction level as a whole, helping distinguish how components in the defined abstraction level interacts with their environment. This  facilitates identifying potential interface issues.  It clarifies the boundaries of the abstraction level and serves as the first high level view of the abstraction level for the producers and consumers involved. This is crucial to building a complete understanding of the system interactions and is thus required to develop the system with quality.  Understanding a context diagram is an essential skill for development and quality engineers alike. 

There are many possible ways to represent a context diagram.  Our style is a grey box approach that uses interfaces and main use cases through the system to capture the desired information. Interfaces may be of two types, external to the abstraction level and internal.  The external interface binds the system and provides a sense on how the abstraction level interacts with the rest of the solution. The internal interfaces gives a sense of how the abstraction level is implemented. The descriptions avoids details intentionally as we only want to convey a sense of the overall boundaries of the abstraction level.  The pictorial used to represent the information is not essential as long as boundaries, interfaces and use cases are conveyed clearly and with minimal details.  The following set of exercises is designed to help clarify the concept of a context diagram.

\begin{exercise}
\label{man}
We revisit the concept of a file on a Unix like file system.  The basic requirement for users to be able to read the content of a file as well as write new data to a file.  The Linux man page details the open() file operation (see \href{https://man7.org/linux/man-pages/man2/open.2.html}{open man page}).  List as many file operations as possible from the Linux man page.
\end{exercise}

\begin{exercise}
The Linux kernel implements file handling.  For the Linux kernel or in our language at the abstraction level in which the Linux kernel implements file handling, the file man page interfaces found in the previous exercise are:
\begin{enumerate}
\item External to the kernel level handling the file handling.
\item Internal to the kernel level handling the file handling.
\item Both internal and external to the kernel level handling the file.
\end{enumerate}
\end{exercise}

\begin{exercise}
The Linux kernel handling the file as follows. First it obtains the current process that is executing the file operations.  It then locates the file.  Once the file is located and provided that the process has appropriate permissions, the file operations are executed and results are returned to the process that invoked the operation.  Suggest use cases that the kernel implements. 
\begin{enumerate}
\item Locate a file and then write to it if the process has permission.
\item Locate a file and then read it thus retrieving its data if the process has permission.
\item Locate a file and obtain is meta data if the process has permission.
\end{enumerate}

Cross check your suggestion above with the many page you have created in the previous exercise()\ref{man}.
\end{exercise}

\begin{exercise}
Suggest internal kernel interfaces that will serve to implement handling of the file.  Choose the most correct option below.
\begin{enumerate}
\item Located a file, read and write to a file
\item Locate a file, read a file, write to a file, obtain the meta properties of a file
\item Locate a file, determine if a process has a permission to a file, read and write a file, obtain the file meta data
\end{enumerate}
\end{exercise}


\begin{exercise}
Create a pictorial representation of the abstraction level of handling a file at the kernel level.  One suggestion follows (see \ref{fig:contextDiagram} below).  A box represents the kernel level abstraction levels.  Lines on the edges represents each the external interfaces.  Inside interfaces are represented as box within the kernel box.  If you follow the above suggestion how would you represent scenario?  Choose the best answer below. 
\begin{enumerate}
\item By lines that connect internal boxes representing internal kernel interfaces.
\item By lines that connect and external interface on the kernel box with internal boxes representing internal kernel interfaces.
\item It is not possible to represent scenarios using this graphic representation.
\end{enumerate}
\end{exercise}

\begin{figure}[h!]
\centering
\includegraphics[scale=0.54]{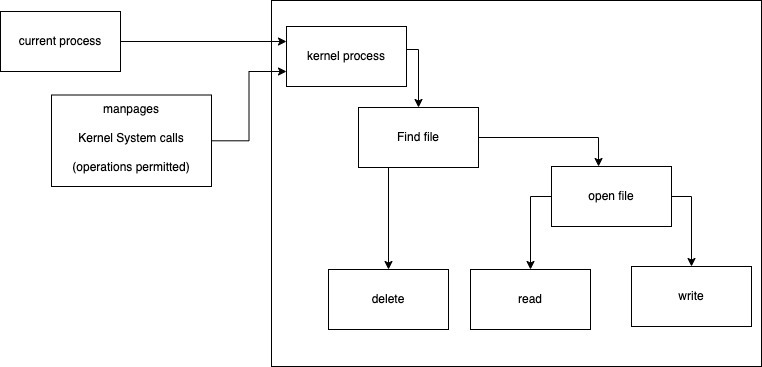}
\caption{File handling context diagram}
\label{fig:contextDiagram}
\end{figure}

Once an overview of the system is created it is time to provide the necessary details that define the uses cases and interfaces.  We deep dive into how this is done next.

\newpage

\subsection{Use cases}

The overview and context work described in the previous section identified the main use cases and interfaces of the current abstraction level.  Whether we first deep dive into the definition of the interfaces and their associated contracts or the use cases is a matter of convenience.  In order to keep the description clean and easy to follow with first deep dive on the definition and use case and deal with interfaces and contracts in the next section.

A Use Case is a description of how a user, sometimes also referred to as actor or persona, uses the solution to achieve their goal.  The goal of the use case typically represents the functional reason for using the solution or requirements of the current abstraction level should meet.  Thus, the goal of the use case should be described in declarative language.  For example, a use case that has the goal of writing new data to a file. Note that our description of the goal is indeed declarative; we are not interested in how the data is added to the file but we declare that this is the desired result.  Stakeholders of the use case are team members that are effected by the use case, e.g., testers, developers, performance stakeholders and  security stakeholders.      

But why are we defining use cases in details?  Defining use cases is a good way to clarify requirements for a given decomposition level.  For example, while defining the use case that implements the writing of data to a file questions like whether or not anyone can write to a file, what happens if the file exists and can you write to any file type are likely to arise.   Use cases also help bridge the gap between the declarative language of the requirements and the imperative language needed to better understand the requirements and how they will be implemented.  The detailed definition of the use cases will indicate if the interfaces defined for this abstraction level are easy to use and make it possible to achieve the goals of the use case.  

To get the full benefit of use case definition, one should elaborate on the interface definitions up to the level of their prototype (input and output specification) and co tracts as discussed in the next section.  At that point the use case naturally transit into a test.  The obtained test is used in the ATDD process to get the most benefit of the use case definition.  The other side of this observation is that if the use case does not transit to a test the design is still ambiguous and not will defined.  Note that one use case may typically translate to several tests.  For example writing new data to a file can translate to a test that writes to a regular file and another test that writes to a socket. Having said that one should be careful not to create too many use cases or make them to detailed and include implementation details.  Use cases are aimed at describing the basic design of the abstraction level and how it achieves the requirements.  It is there to facilitate the communication between the stakeholders.   In order to do that it should clean and simple.  Variations of the basic use case can be mentioned in brief and then then translate to a full blown test in the ATDD level. 

\begin{exercise}
The following set of use cases are defined for file handling:
\begin{enumerate}
\item open a file for read with permission; read the file; close the file
\item open a file for read without permission; read fails
\item open a file for read with permission but the file does not exist; read the file; close the file
\item open a file for write with permission; write the file; close the file
\item open a file for write without permission; write fails
\item open a file for write with permission but the file does not exist; read the file; close the file
\end{enumerate}

Create two use cases out of the above use cases to ease communication between the stakeholders and otherwise improve the description of the use cases below to facilitate the validation of the abstraction level.  Choose the best answer below.

\begin{enumerate}
\item It is not possible to do that as these are different uses cases and all of them needs to be reviewed.
\item \label{two}One use case is opening a file for read the other is opening a file for write. each of the use cases have several variations deepening if the user has permission, the file type and whether it exists or not.
\item Define as in \ref{two} and add a use case that opens the file for read and write simultaneously.   All of the variations defined in \ref{two} apply.
\end{enumerate}

\end{exercise}

\subsection{Interfaces and their contracts}

Use cases are realised through a sequence of interface calls.  Interfaces have an input and an output.  To apply ATTD we need to specify the input and the output of the interface.  At that point the use case can be translated into a test and added to the tests used as an acceptance condition for a development sprint. 

\begin{exercise}
Consider the \href{https://man7.org/linux/man-pages/man2/open.2.html}{man page of the open file API} and determine its inputs and output.  Choose the best description below of the open API input output relation.  

\begin{enumerate}
\item The input includes the file name, the intended file operation and permissions.
\item The The input includes the file name, the intended file operation and permissions.  The output provides the file handle.
\item The The input includes the file name, the intended file operation and permissions.  The output provides the file handle if the open succeeded, an error code otherwise.
\end{enumerate}

\end{exercise}

The use case highlights the relations between interfaces.  In addition, relation between the interfaces can be made more precise using contracts.  For example, if a file open operations succeeds, a file handle is returned.  The file handle should then be used when calling the interfaces that read, write and close the file.  The use case highlights the file handle relation; we can see that a file handle is obtained from the open interface and used when reading, writing and closing the file.  To make the definition of the relation more precise we state that we can state that upon $(fd > 0)~ or~ (fd == -1)$.  $fd$ is the output return value obtained from the open API.  It is either $-1$ indicating a failure or greater than zero indicating a valid file descriptor.  Contracts come in three flavors namely, precondition, precondition and in variants.  The file descriptor example above is a post condition and applies after the API is called.  A precondition is a condition that applies before the API is called.  For example, a write operation to a file requires a valid file descriptor as a precondition. An invariant is a condition that always apply.  For example, the number of entries in a cache that are allocated and the number of entries that are not allocated should add up to the entire size of the cache.  As contracts are logical conditions they can be added to the tests and to the code being implemented thus implementing the preference of code over documentation in yet another way.

\begin{exercise}
what is wrong with the next code snippet?   
$fd = open("myFile", O_RDONLY); n = read(fd, buf, 10);$
Assume $buf$ is allocated enough memory.  Hint - think about the post condition discussed in the proceeding paragraph.
   
\end{exercise}

\subsection{Bringing the stakeholders together to review the results}
\label{reviews}

There are two main Critical Thinking activities that helps find problems in our solution, namely testing to break the system and reviews to make the system stronger upfront.  Reviews are the oldest trick in the book of software engineering.  Here we discuss a specific style of reviews that helps validate an abstraction level just defined.  The review activity brings together the stakeholders in a last team effort to validate a sprint before the ATTD process is invoked.  In this stage we may have, user stories, use cases and contracts, a context diagram, some verbal requirements and a positive and negative glossary defined for the abstraction level.  

One of the "trade secrets" of reviews is comparing two artifacts.  A classical example of the application of that principle is duplicate code.  If you identify similar code segments in different code location compare them!  Many time careful comparison of the two code segments will yield a problem.  Another example of the application of this principle is the identification that a design pattern is being implemented.  If for example the decorator design pattern is being implemented, find some standard implementation of the design pattern on the internet and compare it to the current implementation being reviewed.  Again you are likely  to find problems.  To apply the comparing principle we would like to compare the artifact created above; they should all be consistent with each other.  

Naively attempting to compare the above list of artifacts is tedious.  As we have six artifacts we have $36$ ways to compare them one with the other.   Instead we use another approach.  We use paraphrasing.  The team get together and one of the team members provides and overview and walks every one through the uses cases.  Review participants are assigned an artifact.  For example, one of the reviewers pays attention to the positive and negative glossary.  As the use  cases are being describes the reviewer attempts to break it by finding that the wrong terminology or undefined terminology is used.  Another reviewers focuses on contracts, are the user cases adhering to the contracts?  Yet another reviewer is focusing on the interface definitions and checks if the interfaces are used correctly by the use case.  In this way, comparison between the different artifacts is achieved quickly and efficiently.      

But how do we choose the team that participates in the review?  Reviewers should include the customers of the abstraction levels and their developers.  In addition, reviewers should be chosen based on perspectives they are concerned about.  Also, if our solution is using other APIs it is best to have the owners of the APIs or someone familiar with the APIs semantic present and the review to help validate that we are using the APIs correctly. 

\begin{exercise}
An abstraction level obtains the sum, sum of square and average of a set of numbers stored in persistent storage.  In order to do that a file (define as a sequence of characters) is read, an existing parsing utility is invoked to obtain and array of numbers and a new utility is developed to calculate the sum, sum of squares and average.   Note that the set of numbers can not fit in memory at the same time so the calculation needs to be preformed in chunks that fit in memory.  Specify the existing APIs that are used to implement this abstraction level. 
\begin{enumerate}
\item File operations such as open and read
\item File operations such as open and read, a parsing of a sequence of characters to numbers utility
\item File operations such as open and read, a parsing of a sequence of characters to numbers utility, a utility that calculates the sum, sum of sequences and average 
\item Only 1 and 2 above is correct
\end{enumerate}
\end{exercise}

For more details on how to bring the producers and consumers of a decomposition level to review its definition see \cite{8411768}.

\chapter{Agile project management for product level solutions}

When delivering a product in Agile, there are two important lists that need to be followed and tracked to closure. One is the features list, i.e. the features of the product that need to be developed and the other is the list of activities that lead to the features and the product being built and delivered.  The first is commonly known as the Program Increment that is to be delivered by the sprint activities. The second is the tracking of work done or the sprint activities themselves. While the first is tracked as the product backlog; it is often muddied with lists of activities and items from the work list  or task list that  gets created as the team meets and reviews the work to be done in the daily scrums. As a result, long lists are often created. Further these lists tend to lead to more work being defined and the outcomes from ticking off activities in the list is not always directly translated into the features that are to be produced.  So the work that is often captured and tracked in these lists does not clearly indicate their direct contribution to feature realisation.  A lot of work keeps getting done, and keeps getting added to the list and it is easy to get lost in the work that does not necessarily and directly line up to building a concrete and useful deliverable.  These activities are enablers and should not be confused with the product features being delivered by the team. We need to guard against forgetting what we are delivering vs what we are doing to deliver features and unless we have a very skillful scrum leader, one could get lost in a rabbit hole.


As an example, to develop a design you need inputs from the business, so you must set up meetings with the business team. There many be multiple business teams you have to meet and there can be multiple meetings with each. All these meetings are tasks that need to be tracked as it impacts your ability to get the required inputs to design your product.  But, they are just means to an end and therefore marking off the tasks of 'meeting the business' as done, does not complete the design. It is therefore useful to have a hierarchy of tasks that are associated yet distinct to allow the scrum leader and the team to distinguish between the work and work products. 

A clearly maintained hierarchy enables an accurate and timely view of the progress towards the realization of the features being implemented as well as the effort and duration needed to make it real. In a non-product solution the muddied list might still work when you have single scrum teams contributing to the Agile Release Train and there is no cross-linking with other work streams for the deploy-able work-product / output. Since you do not always know for sure upfront how your product can grow release on release, it is a best practice to always assume that the Agile Release Train has multiple sprints and multiple sprint teams and establish a process upfront to track work products and the work items required to  achieve the work products distinctly. This leads to another topic of the definition of done and the inherent need to establish the definition of done with quality metrics based on quantitative measures. 

\begin{exercise}
The following task list is further refined to additional sub tasks. The original task list includes two requirements, namely:

\begin{enumerate}
\item Add a new file type to the file system that contains audio
\item Reading the audio file system can result in playing the audio
\end{enumerate}

After a scrum meeting the above list is updated as follows.

\begin{enumerate}
\item Add a new file type to Linux mount points that contains audio
\item Add a new file type to Windows mount points that contain audio
\item Obtain an conference room for the weekly team meeting
\item Determine the list of mount points audio files should be supported on
\item Meet Sam who is an expert in networking to see if there are any new security issues with our file system implementation
\item Implement the playing of the audio when reading a file
\end{enumerate}

Which of the following statements are the most correct statement.

\begin{itemize}
\item Obtaining a conference room is a task that muddies the list
\item Obtaining a conference room and the security concern tasks are tasks that muddy the list.  In addition, the different mount point tasks are pointing to the original audio file requirement thus creating an implicit hierarchy which is best made explicit
\item Obtaining a conference room and the security concern tasks are tasks that muddy the list
\item All of the above tasks are valid and the list is the right place to have them. 
\end{itemize}

\end{exercise}

Note: As we bring in rigour to tasks and work product tracking, one might think that we are falling back to standard project management techniques to establish the output being produced and elaborate it with the tasks to be done along with inputs and dependencies. It is not the case, as the process of defining the work required to achieve an outcome is democratised and left of the owner of the work-product to define. It is not dictated nor a pre-determined list of tasks specified against against a hard timeline and structure.  It is left to the owner responsible for producing the deliverable to define how they would do it and how it impacts the deliverable. Tracking that is not the focus of the work plan. 

TBC - the best practice is to have a task list that are grown and discovered during the construction of the solution but is kept structured and clearly focused on the features we are attempting to implement.

The task list is detailed as a continuous process. It is s discovered and defined during the sprint cycle and fine-tuned as the continuous discovery, continuous design, continuous development and continuous test and continuous deployment process unravels.  Usually a tool is used to help the team define the work product and work / activities and the template for defining them is established for the team to adopt upfront. These task and work management and Agile collaboration tools that are useful to  learn and master. Some examples include Azure DevSecOps, Jira Trello boards, Agility, etc, 

To box the effort, the scope of the activity needs to be well-defined and the scope has to be well-bounded. Therefore grooming features and defining activities for building features is an art and key skill to be developed. Otherwise, you risk failing due to over extending the team and the solution.  The way to keep on track and prevent scope escalation, scope creep and scope divergence is to create well-defined acceptance criteria keeping the end user and the end quality of the functional and non-functional aspects of the system in mind. Refer chapter on skills essential to be successful in Agile in addition to review skills and design thinking.

TBC - Planning QE in Agile for Cloud and edge solutions.
CReating a 3X3 matrix to identify the focus areas 
TBC - pitfall - merging project management with review process and design

\chapter{The role of AI}
\label{AI}

How do we utilize AI to increase the quality of the solution?   We distinguish between three potential applications, namely translation, search and sequence anticipation and analysis and elaborate on each of them below.  Some of the examples we provide implement more than one of the possible applications of AI to achieve their goal.  

Foundation models is an emerging form of ML in which a model is trained on huge amount of data and can be used as the starting point for a variety of additional tasked (sometimes called downstream tasks in the literature).  Any one can use large models for tasks such as code, test and documentation generation.  The challenges and opportunities in integrating foundation models into the QE are great.  On the one hand their usage can boost efficiency and quality but on the other hand a new set of techniques should be applied to get value and avoid pitfalls.  One key such pitfalls is factual/correct generation.  we deep dive into this topic in the last section of this chapter.   

Translation means that one artifact is converted by the AI to another artifact.  In \cite{https://doi.org/10.48550/arxiv.2207.13143} API specifications, namely Jaeger documentation of the APIs, are translated into executing code that tests the APIs.  This can serve to ease the task of creating ATDD acceptance tests once the interfaces of the APIs are defined.  Another translation could cluster an existing set of regression tests in order to minimize 
them\cite{10.1145/2950290.2983957} yet another translates requirements to their associated risk\cite{9007287}.  We envision that many more translations are possible, especially with the introduction of \href{https://en.wikipedia.org/wiki/Wikipedia:Large_language_models}{large language models} new use cases will be explored.  For example, it is possible with state of the art large language models to translate description of a low level programming task to useful code snippets.

Using AI to search is a classical application of AI to testing.  For example genetic algorithms are used to test the solution for security vulnerabilities, e.g., see \cite{Qu_2021}.  AI search has also been applied to non-functional chaos directed test generation \cite{https://doi.org/10.48550/arxiv.2109.02540} and to functional testing of APIs  \cite{https://doi.org/10.48550/arxiv.2207.13143}.  Many time the search is guided by an implicit or explicit coverage model that serves to guide the search.  For example, in the case of security vulnerabilities models of code coverage are applied. 

\begin{exercise}

Your chaos probes are able to take done services.  Your solution is composed of two services, namely and online bookshop and a credit validation service.  In addition, your user case includes the following use cases.
\begin{enumerate}
\item Search for a book, add the book to the cart, validate credential, pay for the book
\item Search for a book, book is not found
\item Search for a book, book is found but not in stock
\end{enumerate}
Suggest a coverage model that will test the above use cases taking into account possible failures:
\begin{enumerate}
\item Run the above use cases many times.
\item Run the above two use cases while in parallel taking the bookshop or the credit services down.
\item Run the above two use case while in parallel taking the bookshop or the credit services down and also while taking both the bookshop and the credit down.
\end{enumerate}

If parts of the bookshop solution can be taken down what parts would you take down to better test the solution?
\begin{enumerate}
\item The inventory sub service.
\item The search sub service.
\item Both the inventory and search sub services.
\end{enumerate}

\end{exercise}

The last application of AI occurs through sequences and analysis.  \href{https://en.wikipedia.org/wiki/Anomaly_detection}{Anomaly detection} has been applied solution logs in order to analyze and debug the system.  Even more exciting large language models with generative capabilities are attempting to anticipate the next program statement a programmer enters with some success.  Anomaly detection has nuances and a range of approaches that apply.  For example, a black Friday activity on the system represents a change in the way the system is used but is not a stable ongoing change but a single abrupt occurrence. Techniques have been developed to identify a change that is stable over time and requires a change in the solution.  They are name drift analysis, e.g., \cite{https://doi.org/10.48550/arxiv.2111.05136}.   

Large language models or more general foundation models and their application is a fast growing area and we expect to update this book to reflect its growth.

\begin{exercise}
Using an interactive large language model (e.g., chatGPT) specify that you desire a code for solving  a quadratic equation and see what you get.  Attempt to run it. Did it run?  Think about tests for your quadratic equation solution by your self.  Then ask the large language model for tests to the quadratic equation solution.  Did you get a similar set and what was the difference?
\end{exercise}

What we have done in the preceding exercise is called prompt engineering but with a twist. We used two results obtained with different prompts to validate the solution.  

We next elaborate on how large language models can be used to improve our quality engineering.  


\section{Quality engineering using large language models}

Large language models (LLM) typically are implemented using a transformer architecture. The transformer architecture is based on the use of an attention mechanism combined with an artificial neural network construct \cite{https://doi.org/10.48550/arxiv.1706.03762}(see \href{https://arxiv.org/abs/1706.03762}{link}).  The important take away for the purpose of QE is that the architecture is able to generate sequences of "tokens" from one domain to another which are somewhat related to "languages".   The fundamental examples being  a translation from one language to another in which the "tokens" are words in the language, but in the context of QE we will encounter many other such translations the large language models can be applied to. 

In the case of a natural language such as English the tokens are words.  Words are put together into sentences.  As this is done, the words have a particular role in the sentence construct.   For example, in the "cat sat on the table", cat is a subject and sat is a verb while "one the table" answers the question of where the cat sat.  We refer to this analysis of the natural language sentence as the syntax of the language.  In order to successfully translate from one natural language to another the LLM needs to implicitly understand their syntax.  Similarly, syntax of other potential translation from one language to another, such as code generation from natural language, should be implicitly identified by the LLM.  In the next exercise we consider potential translations in the context of QE.  

\begin{exercise}
\label{RoleofAI:identifyartefact}
Identify possible artifact created during the software engineering development process that can potentially be translated from one to the other using LLM.   Also identify the tokens and syntax in each case.  Choose the best answer below:
\begin{enumerate}
\item Translate use cases to code.  Uses cases are natural language and code has the programming language syntax and vocabulary (tokens). 
\item Translate use cases to tests.  Use cases are natural language and test has the programming language syntax and vocabulary (tokens). 
\item Translation of code to tests and visa versa.  Code and tests have programming language syntax and vocabulary (tokens).
\item Translation of code or tests to use cases.   Syntax and vocabulary as in the previous answers.
\item Translation of user stories to use cases.  Syntax and vocabulary is the syntax and vocabulary of natural language. 
\item Translation of problems logs to potential solutions.  Syntax and vocabulary is the syntax and vocabulary of natural language. 
\item All of the above and more.  Make a few suggestions of your own.
\item All of the above but this is it - no more potential application exists.  
\end{enumerate}
See \ref{RoleofAI:indentifyartefactSolution} for the solution.
\end{exercise}

Now we look at how the artefacts generated by LLMs can help QE process itself?  The trick would be to generate two things that can be compared for validity and verify them. We could  manually compare two artefacts generated by an LLM. This would be a review best practice. As comparison gives a new perspective of the problem and makes it easier to find issues.  Manual comparison is however limited in the number of items one can inspect due to time constraints.  As a result, one can either produce only a small set that needs to be manually compared or randomly choose items to compare out of a larger population. The second approach is to identify an oracle that can automatically determine if two generated artifacts are consistent.  For example, when generating tests and code from a use case, tests can be run against the code to check if the two generated artifacts are consistent.  Any discrepancy in findings is then inspected to determine whether the code or the test need to be fixed.


\begin{exercise}
You obtain a quadratic equation solution implementation from a LLM.  Remember that the equations reads as $ax^2+bx+c = 0$.  Next you asked the LLM for a complete set of tests and got two tests, namely, $a = 1, b = 1, c = 1$ and  $a = -1, b = 1, c = 1$.   Running the tests against the implementation gave correct results.  Your conclusion is as follows.
\begin{enumerate}
\item This is great.  The solution was generated and tested successfully.  
\item I'm not certain that the solution is correct.
\item I'm not certain that the solution is correct.  I want to cover the case in which $a = 0$ as well as the case in which the discernment is bigger than 0, equal to zero and less than zero. 
\item As in the previous answer, but I also want to check the code coverage my tests have achieved.
\end{enumerate}

\end{exercise}

But why are we so worried about correctness of the LLM generated artifact? LLM are generated on a lot of data.  They are known to generate things that are incorrect.  If we are building an industrial ready solution extra care is needed to make sure that the artifact we are incorporating using LLM to increase quality and speed of the development process does not harm the solution quality.




\appendix
\label{App:AppendixA}


\chapter{Solutions}

\begin{solution}
\label{AgilewithQE:BiZreview:Solution} This is the solution for exercise \ref{AgilewithQE:bizreview}.
Option three is your best bet.
It is a best practice to create a summary of the functionality in natural language and then create a list of possible correct or acceptable inputs along with a list of possibly erroneous or unacceptable inputs for the function and share that with the reviewer. When this is done at design time, it helps to explain your understanding of what  you will need to code, and also establish in-scope and out-of-scope boundaries. Do this also helps the reviewer think through the functionality and bring to the forefront any other aspects that they would want you to consider when developing the code.
\end{solution}

\begin{solution}
\label{avoidLosing:abstractionSolution} This is the solution for exercise \ref{avoidLosing:abstraction}.

The second example enables better abstraction and thereby extension of the solution. 
Try out the following algorithm in a language of your choice and determine, how you will design the necessary abstraction.

Type Check:
    Check if x is an integer type (e.g., int, long) using a type checking function available in your programming language.

Integer Calculation:

    If x is an integer:
        Initialize a variable square to 0.
        Implement a loop that iterates x times:
            Inside the loop, add x to square.
        After the loop, return the value of square.

Float Calculation:

    If x is not an integer (assumed to be a float):
        Return the square of x by multiplying it by itself (x * x).

Error Handling:

    If x is not a valid numerical type (e.g., string, boolean), consider:
        Throwing an exception or returning an error code to indicate invalid input.
\end{solution}

\begin{solution}
    \label{avoidLosing:userstoryrefinement:Solution}
In the exercise\ref{avoidlosing:userstoryrefinement}, the review activity could have been a refinement session conducted before the beginning of the sprint, or it could have been a design review meeting or even an implementation of tests in sprint based on some design decisions that were taken earlier.  Given that we are discussing TDD, the review most likely refers to the  process of design validation. 
\end{solution}

\begin{solution}
    \label{avoidLosing:accessfilesystem:Solution}
    The solution for \ref{AvoidLosing:accessfilesystem} is as follows:

    \begin{enumerate}
    
    \item File Opening  - verify existence, access and control / access and error handling for this scenario
    
        \begin{enumerate}
     
        \item Test 1: Open a file the does not exist. Does it return an error or create an empty file?
        \item Test 2: Open a file that exists with READ Access. Does it succeed?
        \item Test 3: Open a file that exists with WRITE Access. Does it succeed?
        \item Test 4: Open the same file twice in different modes. Does it handle concurrent access gracefully?
        \end{enumerate}

    \item Reading: check boundary conditions and error handling for crossing the same
        \begin{enumerate}
        \item Test 5: Read from an empty file. Does it return an empty buffer or error?
        \item Test 6: Read a portion of a file starting from different offsets. Does it access the correct data?.
        \end{enumerate}
   
    \item Writing: Perform WRITE validations, data persistence and overwrite/ append etc.

    \begin{enumerate}
        \item Test 7: Write data to an empty file. Does it store the data correctly? Does it create the file? 
        \item Test 8: Write data to an existing file and overwrite existing content. Does it update the file as expected?
        \item Test 8: Write data to an existing file and append data to existing content. Does it update the file as expected?
        \item Test 9: Write data larger than the file size. What should the expected result be? 
    \end{enumerate}
     \end{enumerate}
\end{solution}
\begin{solution}
    \label{avoidLosing:AgileManifest:Solution}
    This is the solution for \ref{avoidLosing:AgileManifest}. The approach that best reflects a good understanding and alignment with the Agile Manifest is: 

    \item We wrote several tests that represents our next sprint and implemented the code.  Could you join our review and provide feedback on the TDD implementation?

    The above indicates that the team have embraced the concept of writing tests first to validate the hypothesis of the correctness of the functionality before implementing it as code. They also also been specific in asking what they want you to review. 
    
\end{solution}

\begin{solution}
\label{avoidLosing:interfaceUseageSolution}
Solution of exercise \ref{avoidLosing:interfaceUseage}.  The correct answer is the third answer as one needs to take into account the entire cycle of solution usage and maintenance. 
    
\end{solution}

\begin{solution}
\label{avoidLosing:testautomationSolution}
Solution of exercise \ref{avoidLosing:testautomation}. The correct answer is the third answer because an automated tests needs to tell you if it succeeded or not; i.e. was the test passed or failed. 
\end{solution}

\begin{solution}
    \label{RoleofAI:indentifyartefactSolution}
    The solution for \ref{RoleofAI:identifyartefact} is (7)- All of the above and more.  All the listed use cases are good contenders for using LLMs to help create test artefacts. Can you think of more? How about using LLMS to create test data? Or to create a test sequence from an input set of tests?
\end{solution}

\begin{solution}
\label{CriticalTesting:ATTDReadFile:Solution}
Solution for exercise \ref{CriticalTesting:ATTDReadFile}. All options should be a part of your test suite. All combinations described will allow you to critically test the boundaries of the functionality. It will enable you to validate the completeness of the solution implementation. 

Think about how you can avoid duplicating tests from the set above and still cover all the aspects called out.
\end{solution}

 


\chapter{Glossary}

We elaborate here, a list of terms used in the text to establish the common meaning and interpretations that is used throughout the book 


\begin{itemize}
\item Continuous Integration
\item Test First is a methodology of developing  
\item API - Application Programming Interface
\item Domain Driven Design
\item Software-as-a-Service
\item Infrastructure-as-a-Service
\item User stories
\item Domain Driven Design
\item edge
\item Abstraction level  — <this is specific for this book>
\item Waterfall Model <is it worth explaining this or do we assume any reader would/should know already?>
\item AI <probably a common enough term now that it does not need to be listed>
\item full-stack - “Every component within the delivery from database to delivery medium” <or something like that>
\item Design Thinking - Link to chapter 3?
\item “smell check” 
\item ATDD
\item BDD
\item ATTD
\end{itemize}

\printbibliography

@INPROCEEDINGS{8411768,
  author={Route, Saritha and Pendela, Sudheer},
  booktitle={2018 IEEE International Conference on Software Testing, Verification and Validation Workshops (ICSTW)}, 
  title={Combinatorial Test Design – A Smarter Way to Connect with the Business}, 
  year={2018},
  volume={},
  number={},
  pages={306-307},
  doi={10.1109/ICSTW.2018.00065}}

@misc{https://doi.org/10.48550/arxiv.1706.03762,
  doi = {10.48550/ARXIV.1706.03762},
  
  url = {https://arxiv.org/abs/1706.03762},
  
  author = {Vaswani, Ashish and Shazeer, Noam and Parmar, Niki and Uszkoreit, Jakob and Jones, Llion and Gomez, Aidan N. and Kaiser, Lukasz and Polosukhin, Illia},
  
  title = {Attention Is All You Need},
  
  publisher = {arXiv},
  
  year = {2017},
  
  copyright = {arXiv.org perpetual, non-exclusive license}
}

@INPROCEEDINGS{9007287,
  author={Sundararajan, Mukundan and Srikrishnan, Priti and Nayak, Kiran},
  booktitle={2018 3rd International Conference on Contemporary Computing and Informatics (IC3I)}, 
  title={Requirements Complexity Definition and Classification using Natural Language Processing}, 
  year={2018},
  volume={},
  number={},
  pages={76-80},
  doi={10.1109/IC3I44769.2018.9007287}}

@misc{https://doi.org/10.48550/arxiv.2207.13143,
  doi = {10.48550/ARXIV.2207.13143},
  
  url = {https://arxiv.org/abs/2207.13143},
  
  author = {Farchi, Eitan and Prakash, Krithika and Sokhin, Vitali},
  
  title = {Random Test Generation of Application Programming Interfaces},
  
  publisher = {arXiv},
  
  year = {2022},
  
  copyright = {Creative Commons Attribution 4.0 International}
}

@misc{https://doi.org/10.48550/arxiv.2109.02540,
  doi = {10.48550/ARXIV.2109.02540},
  
  url = {https://arxiv.org/abs/2109.02540},
  
  author = {Ackerman, Samuel and Choudhury, Sanjib and Desai, Nirmit and Farchi, Eitan and Gisolfi, Dan and Hicks, Andrew and Route, Saritha and Saha, Diptikalyan},
  
  title = {Towards API Testing Across Cloud and Edge},
  
  publisher = {arXiv},
  
  year = {2021},
  
  copyright = {Creative Commons Attribution Non Commercial Share Alike 4.0 International}
}

@misc{https://doi.org/10.48550/arxiv.2111.05136,
  doi = {10.48550/ARXIV.2111.05136},
  
  url = {https://arxiv.org/abs/2111.05136},
  
  author = {Ackerman, Samuel and Dube, Parijat and Farchi, Eitan},
  
  title = {Using sequential drift detection to test the API economy},
  
  publisher = {arXiv},
  
  year = {2021},
  
  copyright = {Creative Commons Attribution 4.0 International}
}

@inproceedings{10.1145/2950290.2983957,
author = {Zalmanovici, Marcel and Raz, Orna and Tzoref-Brill, Rachel},
title = {Cluster-Based Test Suite Functional Analysis},
year = {2016},
isbn = {9781450342186},
publisher = {Association for Computing Machinery},
address = {New York, NY, USA},
url = {https://doi.org/10.1145/2950290.2983957},
doi = {10.1145/2950290.2983957},
booktitle = {Proceedings of the 2016 24th ACM SIGSOFT International Symposium on Foundations of Software Engineering},
pages = {962–967},
numpages = {6},
location = {Seattle, WA, USA},
series = {FSE 2016}
}

@article{Qu_2021,
doi = {10.1088/1742-6596/2078/1/012015},
url = {https://dx.doi.org/10.1088/1742-6596/2078/1/012015},
year = {2021},
month = {nov},
publisher = {IOP Publishing},
volume = {2078},
number = {1},
pages = {012015},
author = {Sheng Qu and Zheng Zhang and Bolin Ma and Yuwen Shao},
title = {Optimization Method of Web Fuzzy Test Cases Based on Genetic Algorithm},
journal = {Journal of Physics: Conference Series}
}
\end{document}